\documentclass[%draft,superscriptaddress,
reprint,amsmath,amssymb,aps,pra,floatfix]{revtex4-1}
\usepackage[margin=1.6cm]{geometry}

\usepackage{xcolor} %for colored text in comments
\usepackage{graphicx}% Include figure files
\usepackage{dcolumn}% Align table columns on decimal point
\usepackage{upgreek}% Allows text micron
\usepackage{bm}% bold math
\usepackage{hyperref}% add hypertext capabilities
\usepackage{floatrow}
\usepackage[mathlines]{lineno}% Enable numbering of text and display math
\usepackage{physics}

\begin{document}
\long\def\/*#1*/{}

\title{Autonomous Quantum State Transfer by Dissipation Engineering}
\author{Chen Wang}
\email{wangc@umass.edu}
\affiliation{Department of Physics, University of Massachusetts, Amherst, Massachusetts 01003, USA}
\author{Jeffrey M.~Gertler}
\affiliation{Department of Physics, University of Massachusetts, Amherst, Massachusetts 01003, USA}

\date{\today}

\begin{abstract}
Quantum state transfer from an information-carrying qubit to a receiving qubit is ubiquitous for quantum information technology.  In a closed quantum system, this task requires precisely-timed control of coherent qubit-qubit interactions that are intrinsically reciprocal.  Here, breaking reciprocity by tailoring dissipation in an open system, we show that it is possible to autonomously transfer a quantum state between stationary qubits without time-dependent control.  We present the general requirements for this directional transfer process, and show that the minimum system dimension for transferring one qubit of information is 3 $\times$ 2 (between one physical qutrit and one physical qubit), plus one auxiliary reservoir.  We propose realistic implementations in present-day superconducting circuit QED experiments, and further propose schemes compatible with long-distance state transfer using impedance-matched dissipation engineering.
\end{abstract}

\maketitle
\section{Introduction}
Dissipation in a quantum system from its coupling with the environment usually causes decoherence, which has been a major roadblock for quantum information technologies.  In recent years, however, it has been increasingly recognized that dissipation from specifically-engineered environment reservoirs~\cite{poyatos_quantum_1996} can be an important resource for quantum information processing (QIP).  Mostly notably, dissipation can drive a quantum system to relax towards a unique non-trivial steady state.  This steady state can be a resource state such as a Bell state~\cite{krauter_entanglement_2011, lin_dissipative_2013,shankar_autonomously_2013} or a multi-particle entangled state~\cite{barreiro_open-system_2011} for subsequent QIP tasks, or itself can be the potential answer to an open problem, such as a sophisticated many-body state~\cite{diehl_quantum_2008, verstraete_quantum_2009,barreiro_open-system_2011} or the output of a quantum computation algorithm~\cite{verstraete_quantum_2009}.  Moreover, dissipation can be designed to create a steady-state manifold spanned by two or more eigenstates.  This allows confinement of quantum states in a logical subspace~\cite{beige_quantum_2000,zanardi_coherent_2014,leghtas_confining_2015} without disrupting the encoded information, paving the way for possible autonomous quantum error correction~\cite{%lidar_quantum_2013, 
ahn_continuous_2002,kerckhoff_designing_2010,kapit_hardware-efficient_2016,reiter_dissipative_2017,cohen_autonomous_2017,albert_pair-cat_2019}.

Development of the dissipation engineering toobox should ultimately enable implementation of arbitrary quantum processes~\cite{shen_quantum_2017}, which are a far greater set of QIP operations than unitary rotations alone.  %Recent demonstrations in dissipative preparation of resource states~\cite{lin_dissipative_2013,shankar_autonomously_2013} and preservation of a quantum manifold~\cite{leghtas_confining_2015} are therefore only the beginning of many dissipative QIP processes to be explored. 
Here, going beyond %recent demonstrations of 
individual state preparation~\cite{krauter_entanglement_2011,lin_dissipative_2013,shankar_autonomously_2013,barreiro_open-system_2011,hacohen-gourgy_cooling_2015,kienzler_quantum_2015} and manifold confinement~\cite{leghtas_confining_2015,touzard_coherent_2018}, we investigate the feasibility of implementing a dynamic manipulation of a quantum manifold using dissipation: autonomous quantum state transfer (AQST).

%^The task of transfering a quantum state between two subsystems is regularly needed in various QIP applications.  
In a closed quantum system, state transfer between stationary subsystems relies on interactions that swap excitations back and forth, which is reciprocal as required by the Hermiticity of the Hamiltonian.  Precisely-timed external control that turns on and off the swapping Hamiltonian at the right moment is therefore essential for state transfer~\cite{bose_quantum_2003}.  If built-in directionality between subsystems is desired, as is the case for minimizing back-actions in a modular quantum computer~\cite{monroe_large-scale_2014,jiang_distributed_2007} or network~\cite{kimble_quantum_2008}, dissipative reservoirs can be used to construct directional transmission channels~\cite{metelmann_nonreciprocal_2015} to form cascaded quantum systems~\cite{carmichael_quantum_1993}.  While directional transmission of traveling modes can be lossless~\cite{kamal_noiseless_2011}, engineerable~\cite{metelmann_nonreciprocal_2015,ranzani_graph-based_2015} and highly valuable for QIP~\cite{stannigel_driven-dissipative_2012, lodahl_chiral_2017} in its own right, stationary modes necessary for storing quantum information are subject to decay if directly coupled to these directional channels~\cite{carmichael_quantum_1993}.  Therefore, quantum state transfer implemented in cascaded systems so far still requires time-dependent control to dynamically couple and decouple storage modes from the reservoir~\cite{cirac_quantum_1997, Ritter_elementary_2012, kurpiers_deterministic_2018,axline_-demand_2018}.

%Movitated by the need for reducing the formidable control overhead in a complex QIP system, we investigate a type of cascaded system for quantum information
Is it possible to build a cascaded system for quantum information, 
where a quantum state is spontaneously fed forward from an upstream qubit ($A$) to a downstream qubit ($B$) with unit fidelity [as in Fig.~1(a)]? In other words, the ``free" evolution of a two-qubit state without time-dependent external control follows:
\begin{align}
\ket{\psi}_A\ket{\text{vac}}_B\rightarrow\ket{\text{vac}}_A\ket{\psi}_B\nonumber\\
\ket{\text{vac}}_A\ket{\psi}_B\rightarrow\ket{\text{vac}}_A\ket{\psi}_B
	\label{eq:transform}
\end{align}
%This non-unitary transformation implements a quantum state transfer without turning on and off an external-controlled Hamiltonian.
Here $\ket{\psi}=\alpha\ket0+\beta\ket1$ is a logical qubit to be transferred. $\alpha$, $\beta$ are normalized complex coefficients.  The ``vacuum" state, $|\textrm{vac}\rangle$, is a pre-defined state void of information, which can be $|0\rangle$, $|1\rangle$ or an additional non-computational state.  Such autonomous quantum state transfer was first considered in 2008~\cite{pinotsi_single_2008} for atomic emitters with double symmetric lambda structures, which has remained the only physical example of AQST and inspired encouraging experimental progress recently~\cite{bechler_passive_2018}. 
We show that AQST can be quite generally achieved by explicitly synthesizing a dissipative process that 1) acts equivalently on different logical states and therefore blind to the encoded information, and 2) establishes directionality by driving the system into a dark state manifold that stores information in $B$.
%that are blind to logical states.  In essence, the quantum system shall be constructed with two orthogonal degrees of freedom: One encodes a logical qubit and is protected from dissipation, and the other encodes location of the information, which interacts with the reservoir to obtain directionality.  Dissipation asymptotically drives the quantum state onto 

This Article is organized as follows: In Section II we discuss the minimum system size for AQST and the basic reservoir engineering strategy for it.  We then incorporate directional traveling modes to support distinct modularization and remote state transfer in Section III.  Next, we present in Section IV a detailed experimental proposal of AQST in superconducting circuit QED with realistic parameters.  In Section V we show proof-of-principle applicability of AQST in more limited physical systems with pure two-level systems with bilinear interactions.  Finally, in Section VI we comment on general conditions for AQST and its connections to autonomous error correction, followed by an outlook.

\section{Minimum system construction}
The first observation we make from Eq.~(\ref{eq:transform}) is that at least one of the two physical subsystems has to contain more than two eigenstates.  
To prove this by contradiction, we suppose $A$ and $B$ are both two-level systems, and let $|\text{vac}\rangle_i=|0\rangle_i$ ($i=A$ or $B$) without loss of generality.  
The open system ($S$) composed of $A$ and $B$ can be considered as part of a larger closed system that includes the environment ($E$) and undergoes unitary evolution.  Any quantum process %acting on the density matrices in Hilbert space $\mathcal{H}_S=\mathcal{H}_A\otimes \mathcal{H}_B$ 
for $S$ can thus be described by a unitary transformation $\hat{U}$ acting on (the state vectors in) an expanded Hilbert space of $\mathcal{H}_A\otimes \mathcal{H}_B\otimes \mathcal{H}_E$, followed by tracing out $E$.  To satisfy Eq.~(\ref{eq:transform}), for any input state vectors of the form $\ket{\psi}_A\ket{0}_B\ket{x}_E$ or $\ket{0}_A\ket{\psi}_B\ket{x}_E$ (where $\ket{x}$ is a state vector in $\mathcal{H}_E$), $\hat{{U}}$ must not entangle $S$ with $E$%(so that the final state remains pure after tracing out $E$)
, therefore $E$ itself undergoes fixed unitary transformation ($\hat{R}_1$ or $\hat{R}_2$):
\begin{align}
\hat{{U}} \big( \alpha|00\rangle+\beta|10\rangle \big)_{S} \ket{x}_E = \big( \alpha|00\rangle+\beta|01\rangle \big)_{S} \hat{R}_1|x\rangle_E \nonumber\\
\hat{{U}} \big( \alpha|00\rangle+\beta|01\rangle \big)_{S} \ket{x}_E = \big( \alpha|00\rangle+\beta|01\rangle \big)_{S} \hat{R}_2|x\rangle_E
	\label{eq:separable}
\end{align}
For an input state $\ket{\phi}=\big(\alpha\ket{00}+\beta\ket{10}+\gamma\ket{01}\big)_{S}\ket{x}_E$, using different linear combinations of Eq.~(\ref{eq:separable}), we get
$\hat{{U}}\ket{\phi}=\big(\alpha\ket{00}+\beta\ket{01}\big) \hat{R_1}\ket{x}+\gamma\ket{01} \hat{R_2}\ket{x}$ and 
$\hat{{U}}\ket{\phi}=\big(\alpha\ket{00}+\gamma\ket{01}\big) \hat{R_2}\ket{x}+\beta\ket{01} \hat{R_1}\ket{x}$
for arbitrary coefficients $\alpha$, $\beta$ and $\gamma$.  This requires $\hat{R}_1=\hat{R}_2$, so
\begin{equation}
\hat{{U}} \big( \alpha|00\rangle+\beta|10\rangle+\gamma|01\rangle \big) |x\rangle = \big[\alpha\ket{00} + (\beta+\gamma)\ket{01} \big] \hat{R}_1\ket{x} \nonumber
%\alpha\ket{00}+\beta\ket{01}+\gamma\ket{10} \rightarrow \alpha\ket{00} + (\beta+\gamma)\ket{01}
%	\label{eq:non-unitary}
\end{equation}
Therefore, $\hat{{U}}$ is a deterministic non-unitary quantum gate, which is not only contradictory to its definition but also forbidden within the framework of linear quantum mechanics~\cite{terashima_nonunitary_2005, abrams_nonlinear_1998}.%: the trace of the density matrix of the system (in $\mathcal{H}_A\otimes \mathcal{H}_B$) is no longer equal to 1 after this transformation.

\begin{figure}[tbp]
    \centering
    \includegraphics[scale=1]{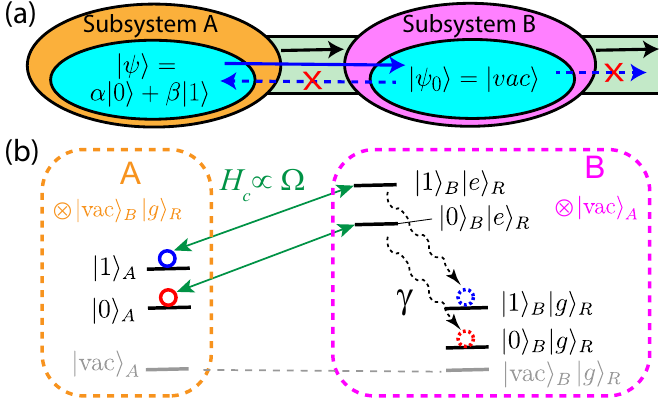}%[width=8.0cm]
    \caption{(a) A conceptual diagram of AQST, where an encoded quantum state $\ket{\psi}$ is spontaneously emitted from a subsystem $A$ and fully absorbed by a subsystem $B$.  This is realized via an symmetric coupling channel which is blind to $\ket{\psi}$. (b) Basic construction of AQST using a reservoir mode $R$. %formed with subsystems $A$, $B$ and a dissipative reservoir $R$.  
The two dashed boxes encloses quantum states with information stored in $A$ (orange) and in $B$ (magenta) respectively.  The two gray levels represent the same global vacuum state, which is never populated if both $A$ and $B$ use non-computational states for vacuum states, as is the case shown.  However, either $A$ or $B$ is allowed to use $\ket{0}$ or $\ket{1}$ as vacuum, in which case the global vacuum will merge with one of the computational states.}
\end{figure}

Now we allow one subsystem to have a non-computational eigenstate as its vacuum state, \textit{i.e.~}$|\text{vac}\rangle_A\equiv|2\rangle_A$.  This ensures orthogonality among the four relevant global eigenstates: The two initial states ($\ket{\sigma,\text{vac}}$) and the two final states ($\ket{\text{vac},\sigma}$) that encode $\sigma=$ 0 or 1. 
Consider a system with Hamiltonian $\hat{H}=0$, starting from an initial state of $|\psi\rangle_A|\text{vac}\rangle_B$, AQST can be achieved by engineering a jump operator (via a Markovian reservoir) of
\begin{equation}
\hat{L}=\sqrt{\kappa}\big[|\text{vac}\rangle_A|0\rangle_B\langle\text{vac}|_B\langle0|_A+|\text{vac}\rangle_A|1\rangle_B\langle\text{vac}|_B\langle1|_A\big]
	\label{eq:3x2jump}
\end{equation}
where $\kappa$ is the jump rate.  This dissipation process explicitly maps the two initial eigenstates onto the two final eigenstates respectively.  The quantum jump $\hat{L}$ will occur once and only once throughout the process. % with an exponential time constant of $1/\kappa$.  
%As can be verified by solving the Lindblad mater equation, the density matrix evolves over time $t$ as $\rho(t)=e^{-\kappa t}|\psi\rangle_A|0\rangle_B\langle0|_B\langle\psi|_A+(1-e^{-\kappa t}) |2\rangle_A|\psi\rangle_B\langle\psi|_B\langle2|_A$.
%\begin{align}
%\rho(t)=e^{-\kappa t}|\psi\rangle_A|0\rangle_B\langle0|_B\langle\psi|_A+(1-e^{-\kappa t}) |2\rangle_A|\psi\rangle_B\langle\psi|_B\langle2|_A
%\end{align}
Although the system evolves as a mixed state, with density matrix evolution $\rho(t)=e^{-\kappa t}|\psi\rangle_A|\text{vac}\rangle_B\langle\text{vac}|_B\langle\psi|_A+(1-e^{-\kappa t}) |\text{vac}\rangle_A|\psi\rangle_B\langle\psi|_B\langle\text{vac}|_A$, at $t\gg1/\kappa$, it exponentially converges to a pure state, and the quantum state is transferred with fidelity arbitrarily close to 1.  %Practically, $t$ in the range of $5/\kappa$ to $10/\kappa$ will be sufficiently long so that infidelity due to incomplete transfer is insignificant compared with other imperfections such as decoherence.  
This process is thus possible for a system dimension as small as 3$\times$2.

%\textit{Dissipation engineering with auxilliary modes}
To engineer the jump operator Eq.~(\ref{eq:3x2jump}), as shown in Fig.~1(b), we introduce a cold auxiliary two-level reservoir $R$ with interaction $\hat{H}_c$ with the system and a simple relaxation process $\hat{L}_r$ (from its excited state $\ket{e}_R$ to its ground state $\ket{g}_R$): 
\begin{align}
	\hat{H}_{c}&=\hbar\Omega \big( \hat{A}_0 \hat{B}_0^\dagger + \hat{A}_1 \hat{B}_1^\dagger \big) \hat{r}^\dagger 
%	|\text{vac}\rangle_A|\sigma\rangle_B\langle\text{vac}|_B\langle\sigma|_A  + c.c.
	\label{eq:couplingH}\\
	\hat{L}_r &= \sqrt{\gamma} \hat{r}
\end{align}
where we define $\hat{A}_\sigma\equiv|\text{vac}\rangle_A\langle\sigma|_A$,  $\hat{B}_\sigma\equiv|\text{vac}\rangle_B\langle\sigma|_B$, and $\hat{r}\equiv|g\rangle_R \langle e|_R$.  % A generic recipe for constructing AQST is shown in Fig.~1(b).  
%Coherent Hamiltonian interactions with equal strengths, are applied to induce transitions from the 
The resulting dynamics is under-damped ($\gamma<4\Omega$) or over-damped ($\gamma>4\Omega$) oscillation between the two initial eigenstates and two intermediate states, but in any case converges to the two final states with a rate of $\text{Re}\big[\gamma-\sqrt{\gamma^2-16\Omega^2}\big]/2$.  The entire process is two-fold degenerate, analogous to optical pumping with a hidden degree of freedom that can be used to encode a qubit.  In the limit of $\Omega\ll\gamma$, the reservoir can be adiabatically eliminated from the Hamiltonian, leaving effectively a jump operator of the form of Eq.~(\ref{eq:3x2jump}) with $\kappa=\kappa_\text{eff}\equiv 4\Omega^2/\gamma$.  

%The two initial eigenstates are driven towards a pair of intermediate states and subsequently decay into the final eigenstates [Fig.~1(b)].  This process is analogous to optical pumping with a hidden degree of freedom that can be used to encode a qubit.  In the limit of $\Omega\ll\gamma$, the reservoir can be adiabatically eliminated from the Hamiltonian, leaving effectively a jump operator of the form of Eq.~(\ref{eq:3x2jump}) with $\kappa=\kappa_\text{eff}\equiv 4\Omega^2/\gamma$.  For arbitrary $\Omega$ and $\gamma$, the system plus ancilla dynamics is described by an underdamped ($\gamma<4\Omega$) or overdamped ($\gamma>4\Omega$) oscillator, but in any case converges to the final state with a rate of $\text{Re}(\gamma-\sqrt{\gamma^2-16\Omega^2})/2$.

%We note that the coupling Hamiltonian of Eq.~(\ref{eq:couplingH}) implies three-body non-local interactions that may appear quite restrictive and challenging to implement in experiments.  From here, we will expand on three key observations that enable many practical avenues to AQST: 1) Non-local interactions can be reduced to local interactions by intermediate communcations modes; 2) There are several physical systems with native support for three-body interactions; 3) Three-body interactions may be reduced to two-body bi-linear interactions using composite degrees of freedom. 

\section{Directional coupling and remote transfer}
The directional nature of AQST discussed here is twofold.  On the one hand, the ultimate effect of information transfer is directional (from $A$ to $B$) by virtue of dissipation, regardless of intermediate dynamics.  On the other hand, one may demand a more stringent form of directionality: the state of $B$ shall have no influence on $A$ on any time scale and regardless of the state of auxiliary modes.  This lack of back-action is often a defining feature of quantum state transfer between remote or distinctive modules, which is not fully achieved in Fig.~1(b) because the underlying connection between $A$ and $B$, as described by a coherent Hamiltonian Eq.~(\ref{eq:couplingH}), remain bidirectional.

The coupling between $A$ and $B$ can be rendered strictly directional when mediated by one or more communication ancilla in a cascaded system setting (Fig.~2).  Communication in a cascaded quantum system~\cite{carmichael_quantum_1993} is necessarily exposed to an information leakage channel due to the open-system nature of directional traveling waves.  For this reason, time-domain waveform shaping has been essential in demonstrations of remote quantum state transfers~\cite{Ritter_elementary_2012, kurpiers_deterministic_2018,axline_-demand_2018}.  Inspired by earlier work of Ref.~\cite{pinotsi_single_2008} and \cite{koshino_implementation_2013}, we show in this section that information leakage can be fully suppressed using ``impedance-matched'' reservoir engineering in the adiabatic limit, thus enabling AQST for long-distance quantum communication.

We consider a scheme with two additional ancilla modes $a$ and $b$ added to the minimal system (Fig.~2): ancilla $a$ locally interacts with $A$ and emits information into a directional traveling wave mode; the receiving ancilla $b$, which is the more essential of the two, locally interacts with $B$ and $R$ in a similar way as in Fig.~1(b).  Each ancilla has a ground state and two excited states preferably from independent excitation modes.  For example, $a$ and $b$ may each be a two-mode cavity, with no photon representing $\ket{\text{vac}}$, a red photon ($\hat{a}_0$ or $\hat{b}_0$) representing $\ket{0}$, or a blue photon ($\hat{a}_1$ or $\hat{b}_1$) representing $\ket{1}$ (or %perhaps more practically for atomic systems, 
using different photon polarization). %The interaction between $a$ and $b$ can be described using the formalism of cascaded quantum system~\cite{Carmichael1993}
The directional channel between $a$ and $b$ can be realized by a chiral waveguide~\cite{lodahl_chiral_2017}, a circulator, or the reservoir engineering scheme of balancing Hamiltonian interactions ($\hat{H}_{ab}$) with dissipative interactions ($\hat{L}_0$ and $\hat{L}_1$)~\cite{metelmann_nonreciprocal_2015}:
\begin{align}
%\hat{H}_{ab}= \hbar\frac{\sqrt{\kappa_a\kappa_b}}{2} \sum_{\sigma=0,1} i[\hat{a}_\sigma^\dagger\hat{b}_\sigma-\hat{a}_\sigma\hat{b}_\sigma^\dagger]\,\,\,\,\,\,\,\,\,\,\,\,\\
\hat{H}_{ab}= \frac{i\hbar}{2}\sqrt{\kappa_a\kappa_b}\big(\hat{a}_0^\dagger\hat{b}_0-\hat{a}_0\hat{b}_0^\dagger+\hat{a}_1^\dagger\hat{b}_1-\hat{a}_1\hat{b}_1^\dagger \big)\,\,\\
\hat{L}_0= \sqrt{\kappa_a}\hat{a}_0+\sqrt{\kappa_b}\hat{b}_0, \,\,\, \hat{L}_1= \sqrt{\kappa_a}\hat{a}_1+\sqrt{\kappa_b}\hat{b}_1
\end{align}
Here $\kappa_a$ and $\kappa_b$ can be understood as the leakage rate of the cavities $a$ and $b$ to the waveguide.  

%The jump operators $\hat{L}_0$ and $\hat{L}_1$ represent an information leakage channel for the quantum state, which is generic to any cascaded quantum system.  For this reason, time-domain waveform shaping has been essential in demonstrations of remote quantum state transfers~\cite{Ritter_elementary_2012, kurpiers_deterministic_2018,axline_-demand_2018}.  Remarkably, with a dissipative subsystem $B$ as the information receiver, information leakage can be suppressed for a fairly flexible range of parameters, allowing for determistic transfer of the quantum state without time-dependent control. 

\begin{figure}[tbp]
    \centering
    \includegraphics[scale=1.00]{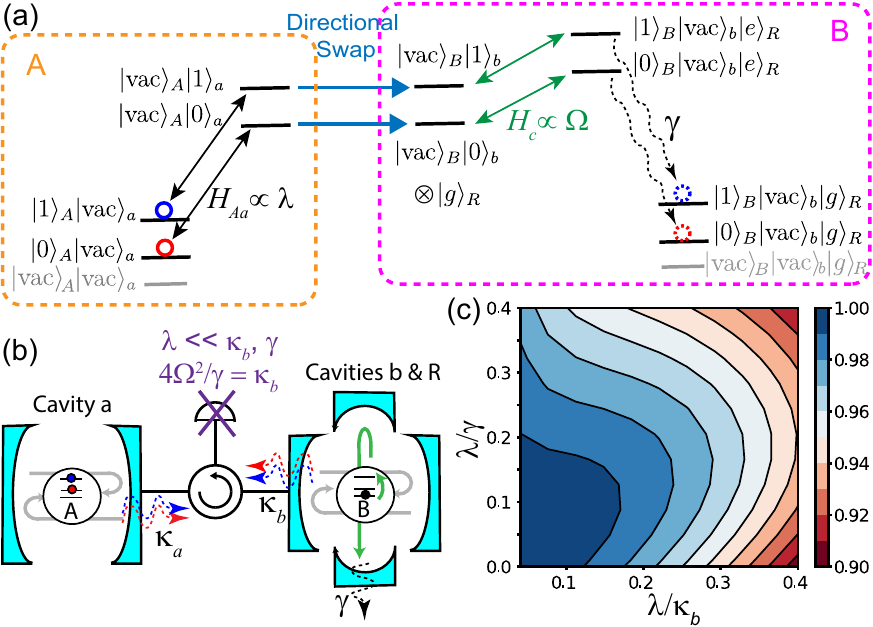}%[width=8.0cm]
    \caption{(a) Energy level diagram and transfer pathway of an AQST scheme between two quantum modules $A$ and $B$ separated by a directional transmission channel.  The corresponding Hamiltonian and dissipation operators are described by Eqs.~(5-10).  (b) A cavity QED representation of this AQST scheme, where photons in red/blue superposition propagate across a waveguide with a circulator, and the receiving cavity $b$ (horizontal) dissipates the incoming photons by down-converting them to a lossy cavity mode R (vertical) while exciting a storage qutrit.  The detector at the third port of the circulator does not ``click'' when the two labeled conditions are met (see text). (c) Numerically calculated intrinsic AQST fidelity as a function of coupling and dissipation rates.  We fix $\kappa_b=4\Omega^2/\gamma$ to satisfy impedance matching condition, and $\kappa_a=\kappa_b$ for convenience.
    }
\end{figure}

The dissipative dynamics of the global system can be described by a stochastic wave function evolving according to the Schrodinger equation $i\hbar\frac{d\ket{\phi(t)}}{dt}=\hat{H}_\text{eff}\ket{\phi(t)}$~\cite{carmichael_quantum_1993, pinotsi_single_2008} with non-Hermitian effective Hamiltonian:
\begin{align}
&\hat{H}_\text{eff}=\hat{H}_{A}+\hat{H}_{ab}+\hat{H}_{c}-\frac{i\hbar}{2}(\hat{L}_0^\dagger \hat{L}_0+\hat{L}_1^\dagger \hat{L}_1+\hat{L}_r^\dagger \hat{L}_r)
	\label{eq:no-jump}\\
&\hat{H}_{A}=\hbar\lambda\big( \hat{a}_0^\dagger \hat{A}_0 +\hat{a}_1^\dagger \hat{A}_1 \big) + c.c. \\
&\hat{H}_{c}=\hbar\Omega\big( \hat{b}_0 \hat{B}_0^\dagger + \hat{b}_1 \hat{B}_1^\dagger \big) \hat{r}^\dagger + c.c. \label{eq:couplingH2}
%\\&\hat{L}_r=\sqrt{\gamma}\hat{r}=\sqrt{\gamma}\ket{g}_R\bra{e}_R
%&\hat{L}_\text{eff}=\kappa_\text{eff} \big[\hat{b}_0\ket{0}_B\bra{\text{vac}}_B + \hat{b}_1\ket{1}_B\bra{\text{vac}}_B\big]
\end{align}
%$\kappa_\text{eff}$ is the effective dissipation rate after adiabatic elimination of the reservoir mode $R$.
The coherent evolution of the stochastic wave function is eventually stopped by jump operators $\hat{L}_r$, $\hat{L}_0$ or $\hat{L}_1$.  Occurrence of $\hat{L}_r$ completes the AQST, but $\hat{L}_0$ or $\hat{L}_1$ collapses the system to the global vacuum, equivalent to a measurement of the quantum state by the environment.  

The occurrence probability of the undesirable $\hat{L}_0$ or $\hat{L}_1$ jump approaches zero when the following two conditions are satisfied:  The first is ``impedance matching''~\cite{pinotsi_single_2008, koshino_implementation_2013}, or the engineered dissipation rate in receiving cavity $b$ matches waveguide coupling: $\kappa_\text{eff}\equiv4\Omega^2/\gamma=\kappa_b$. This ensures full steady-state absorption of incoming signals from the waveguide by $b$.  The second condition is ``adiabaticity'' to minimize reflections of transient signals by $b$.  This requires the incoming signal has a narrow bandwidth compared to the receiving cavity, obtainable by $\lambda\ll\kappa_b, \gamma$ (or alternatively, $\kappa_a\ll\kappa_b, \gamma$).  By numerically solving the system dynamics without considering practical imperfections (such as transmission loss or additional decoherence channels), we found that intrinsic state transfer fidelity of greater than $99\%$ can be achieved for very modest ratios of $\lambda/\kappa_b$ and $\lambda/\gamma$ ($\sim0.15$) [Fig.~2(c)].  We further note $\Omega\ll \gamma$ is not a necessary condition for either impedance matching or adiabaticity.  

To provide more intuition to the state transfer dynamics, we analytically solve the no-jump evolution of $\ket{\phi(t)}$ in the small $\lambda$ limit (Appendix A).  Over a transient period ($t\sim 1/\text{Min}[\kappa_a,\kappa_b,\kappa_\text{eff}]$), $\ket{\phi(t)}$ exponentially converges to a meta-stable state (to first order in $\lambda/\kappa_a$):
\begin{align}
    \ket{\phi_\infty}=&\ket{\psi}_A\ket{\text{vac}}_a\ket{\text{vac}}_b\ket{\text{vac}}_B\ket{g}_R\nonumber\\
    &-(2i\lambda/\kappa_a)\ket{\text{vac}}_A\ket{\psi}_a\ket{\text{vac}}_b\ket{\text{vac}}_B\ket{g}_R\nonumber\\
    &+(2i\lambda/\sqrt{\kappa_a\kappa_b})\ket{\text{vac}}_A\ket{\text{vac}}_a\ket{\psi}_b\ket{\text{vac}}_B\ket{g}_R\nonumber\\
    &+(2\lambda/\sqrt{\kappa_a\gamma})\ket{\text{vac}}_A\ket{\text{vac}}_a\ket{\text{vac}}_b\ket{\psi}_B\ket{e}_R
    \label{eq:wavefunction}
\end{align}
This is a dark state to the waveguide jump operators as %\textit{i.e.~}
$\hat{L}_0 \ket{\phi_\infty}=\hat{L}_1\ket{\phi_\infty}=0$.  There is a non-zero probability rate, $\bra{\phi_\infty}\hat{L}_r^\dagger\hat{L}_r\ket{\phi_\infty}=4\lambda^2/\kappa_a$, for the AQST jump $\hat{L}_r$ to collapse the wave function to complete the transfer.  The infidelity $1-\mathcal{F}$ due to information leakage during the transient stage is given by a time-integral of the probabilities of $\hat{L}_0$ and $\hat{L}_1$, which shows a simple quadratic scaling when we further take the limit of $\Omega \ll \gamma$:
\begin{equation}
1-\mathcal{F} = \int_{t=0}^{\infty} \sum_{\sigma=0}^1 \langle\phi|\hat{L}_\sigma^\dagger\hat{L}_\sigma|\phi\rangle \mathrm{d}t=\frac{\lambda^2}{2\kappa_b(\kappa_a+2\kappa_b)}
\end{equation}
In practice, one may choose $\lambda$ to balance faster transfer speed and smaller infidelity from transient reflections.

\section{Implementation in circuit QED}
Implementation of the AQST, whether or not mediated by communication ancillae, relies on the ability to engineer Hamiltonian of the form $\hat{H}_c$ [Eq.~(\ref{eq:couplingH}) or (\ref{eq:couplingH2})], which is a three-particle interaction analogous to parametric down-conversion or three-wave mixing.  Some physical systems naturally provide such three-particle interactions, such as the atomic lambda emitter that enables single-photon Raman scattering.  Adding a two-fold degeneracy to such a process (\text{i.e.} via a spin degree of freedom) allows information to be carried in the conversion~\cite{pinotsi_single_2008, bechler_passive_2018}.  Another example is superconducting circuit QED~\cite{wallraff_strong_2004}, where the Josephson four-wave mixing Hamiltonian can be used to engineer the three-particle interaction.  In this section, we show that AQST can be demonstrated in readily available circuit QED experimental setup with modest parameters.  

\begin{figure}[tbp]
    \centering
    \includegraphics[scale=.67]{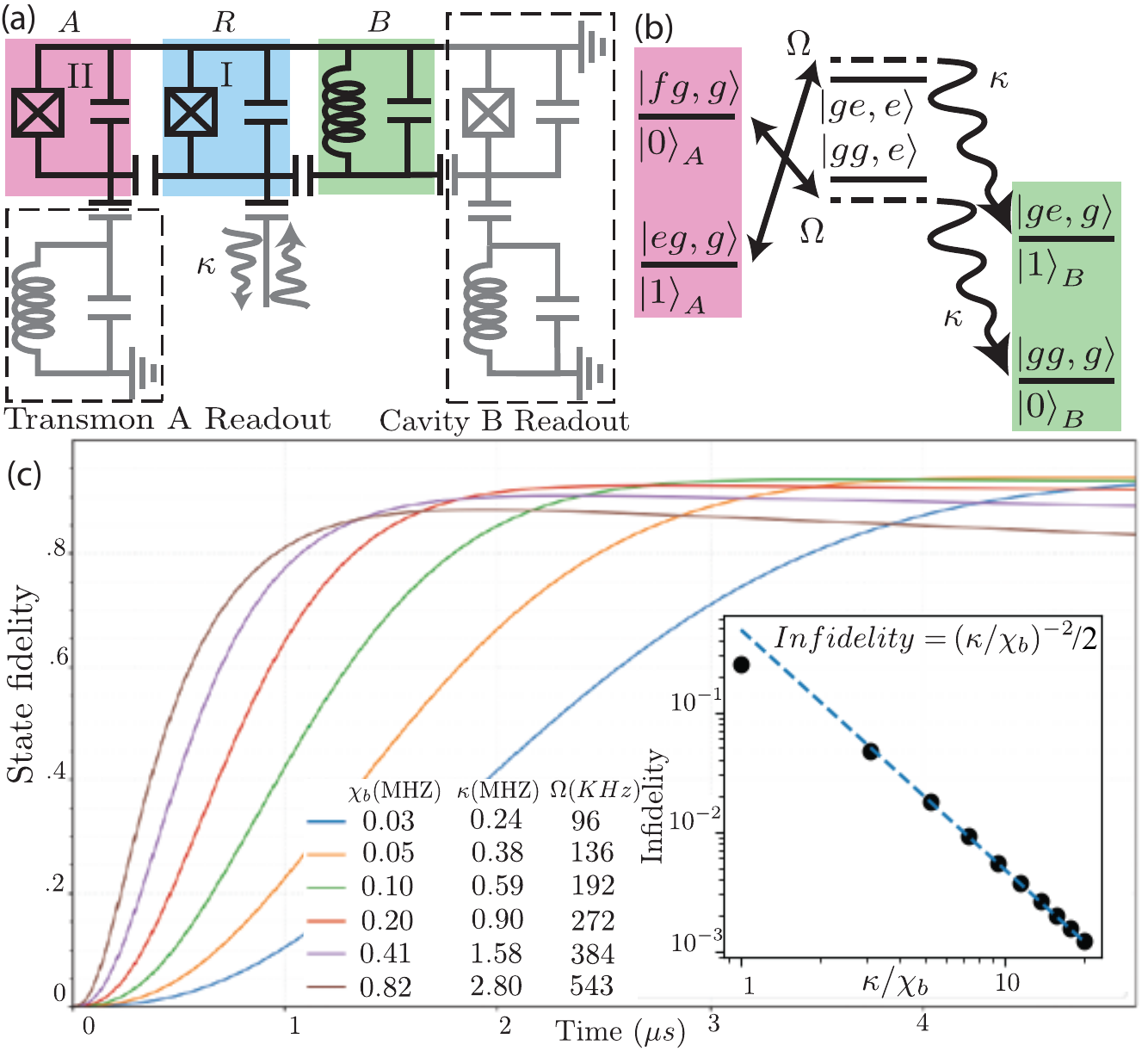}%[width=8.0cm]
    \caption{Implementation of AQST in circuit QED. (a) Effective circuit diagram of a transmon qutrit $A$ (with Josephson junction II), a storage cavity $B$, a reservoir transmon $R$ (with junction I), and auxiliary elements for state preparation and readout.  (b) Energy level diagram that shows the state transfer paths, similar to Fig.~1(b). %The quantum state initially encoded in $A$ is driven to a pair of virtual states by slightly-detuned Rabi drives with equal rate $\Omega$ (straight arrows), and subsequently decay to the final states in $B$ by reservoir dissipation (twisted arrows).  
    (c) Numerical results of transferring an equator state, including decoherence, showing fidelity of instantaneous state $\rho$ against target state $\ket{\phi}$, $\mathcal{F}=\big[\text{Tr}(\sqrt{\rho}|\phi\rangle\langle\phi|\sqrt{\rho})^{1/2}\big]^2$, as a function of time during the transfer.  Different color curves are simulated for different $\chi_b$ and their corresponding achievable $\Omega$ and optimal $\kappa$, all divided by $2\pi$.  Inset shows the ideal-case infidelity due to rotation of the virtual states in drive frame which scales as $(\kappa$/$\chi_b)^{-2}$. 
    }
\end{figure}

Figure 3(a) shows a superconducting circuit that realizes the minimal model of AQST composed of one qubit and one qutrit, which can also be employed as a waveguide receiving node in a larger AQST scheme.  We consider a transmon qutrit~\cite{koch_charge-insensitive_2007} %the most widely-used type of superconducting qubits, 
and a superconducting cavity as the subsystems $A$ and $B$, simultaneously coupled to another transmon qubit $R$ acting as a reservoir with decay rate of $\kappa$.  %The transmons are anharmonic LC oscillators, each made of a Josephson inductance shunted by a capacitor~\cite{koch_charge-insensitive_2007}. 
We only access the lowest three levels ($\ket g, \ket e, \ket f$) of $A$ and the lowest two levels of both $R$ and the cavity $B$.  Computational and non-computational states are defined as
%$\ket 0_A \equiv \ket f_A, \ket 1_A \equiv \ket e_A, \ket{\text{vac}}_A \equiv \ket g_A,
%\ket 0_B \equiv \ket g_B, \ket 1_B \equiv \ket e_B, \ket{\text{vac}}_B \equiv \ket g_B$.
\begin{align}
\ket 0_A \equiv \ket e_A, \ket 1_A \equiv \ket f_A, \ket{\text{vac}}_A \equiv \ket g_A\nonumber\\
\ket 0_B \equiv \ket e_B, \ket 1_B \equiv \ket g_B, \ket{\text{vac}}_B \equiv \ket g_B
\end{align}
AQST is realized by two continuous-wave off-resonant pumps to induce the $\ket{eg,g}\rightarrow\ket{ge,e}$ and $\ket{fg,g}\rightarrow\ket{gg,e}$ transitions (with sequential indices of $A$, $B$ and $R$ omitted) with equal Rabi rates $\Omega$ [(Fig.~3(b)]. 

The cQED Hamiltonian incorporating two off-resonant drives applied to the reservoir mode with normalized amplitudes $\xi_1$ and $\xi_2$ can be written as~\cite{leghtas_confining_2015}
\begin{align}
\hat{H}&=\hbar\omega_A \hat{a}^\dagger \hat{a}+\hbar\omega_B \hat{b}^\dagger \hat{b} +\hbar\omega_R \hat{r}^\dagger \hat{r} \nonumber\\
&-\sum_{i=I, II}\frac{E_{Ji}}{24}\big[\Phi_{Ai}\hat{a}
+\Phi_{Bi}\hat{b}
+\Phi_{Ri}\big(\hat{r}+\tilde{\xi}_1
+\tilde{\xi}_2 \big)+ h.c.\big]^4
\label{eq:main_4order}
\end{align}
where $\hat{a}^\dagger$, $\hat{b}^\dagger$ and $\hat{r}^\dagger$ are creation operators of oscillator modes $A$, $B$ and $R$. $E_{Ji}$ is the Josephson inductance of junction $i$ (= I or II).  $\Phi_{Xi}$ is the zero-point flux fluctuation of mode $X$ (=$A$, $B$, or $R$) across junction $i$.  %We use Junction II as a resource of three-body nonlinear coupling with relatively large $\Phi_{A\text{II}}\Phi_{\text{II}\beta}\Phi_{R\beta}$ product.  Junction $\alpha$ is solely for the purpose of providing anharmonicity to $A$, whose coupling to $B$ is negligible (\textit{i.e.~}$\Phi_{B\alpha}\approx 0$).  
Here we have taken the cosine expansion of Josephson energy to the 4$^{th}$ order, and the drive terms have been absorbed into the Josephson nonlinearity after a displacement transformation (see Appendix B). 
The frequencies of the drive tones, $\omega_1$ and $\omega_2$, are chosen close to
the two aforementioned transitions (with small detunings $\delta_1$ and $\delta_2$), and near-stationary 4$^{th}$-order terms of the form $\xi_1\hat{a}\hat{b}^\dagger\hat{r}^\dagger$ and $\xi_2\hat{a}^2\hat{r}^\dagger$ $(+h.~c.)$ emerge as a result of four-wave mixing.  Under the rotating wave approximation (RWA), the Hamiltonian in the reference frame of the drives is
\begin{align}
\hat{H}_{\text{rot}}=&\hbar\delta_1 |ge,e\rangle \langle ge,e|+\hbar\delta_2 |gg,e\rangle\langle gg,e| \nonumber\\
+& \hbar \Omega_1|ge,e\rangle\langle eg,g| + \hbar \Omega_2 |gg,e\rangle\langle fg,g| + h.~c.
	\label{eq:H_rot}
\end{align}
where $\chi_b\approx E_{\text{JI}}\Phi_{B\text{I}}^2\Phi_{R\text{I}}^2/\hbar$ is the dispersive shift between $B$ and $R$.  The Rabi drive rates are $\Omega_1 = \sum_i E_{Ji} \xi_1 \Phi_{Ai}\Phi_{Bi}\Phi_{Ri}^2 /\hbar$ and $\Omega_2 = \sum_i E_{Ji} \xi_2 \Phi_{Ai}^2\Phi_{Ri}^2/2\hbar$.  
To implement the protocol, $\xi_1$ and $\xi_2$ are chosen to satisfy $\Omega_1=\Omega_2\equiv \Omega$.

The reservoir loss operator, $\hat{L}=\sqrt{\kappa}\hat{r}$, for relevant states in the Heisenberg picture of the drive frame is
\begin{equation}
\hat{L}_{\text{rot}}=\sqrt{\kappa}\big[|gg,g\rangle\langle gg,e|+ e^{i(\delta_2-\delta_1-\chi_b
)t}|eg,g\rangle\langle eg,e|\big]
\end{equation}
The-time dependent phase factor in $\hat{L}_{\text{rot}}$ indicates a dephasing effect due to the energy difference of the reservoir emission for logical $\ket{0}$ versus $\ket{1}$.  To eliminate this error, we choose detunings $\delta_1=-\chi_b/2$ and $\delta_2=\chi_b/2$ to make $\hat{L}_\text{rot}$ stationary.  Effectively, we drive the two sets of $\Lambda$-transitions through nearby virtual states to compensate for the dispersive shift of the real states.  %These symmetrically chosen detunings also ensure equal rates %($=\sqrt{\Omega^2+(\chi_b/2)^2}$) for the two detuned Rabi drives.  
The different rotation axes of the two detuned Rabi drives leads to a rotation of the logical state, but the resulted infidelity has a favorable scaling of $\chi_b^2/2\kappa^2$ and independent of $\Omega$ as we found analytically (Appendix C) and numerically [Fig.~3(c) inset]. 

We performed master equation simulation under the RWA including the 12 basis states of the $A+B+R$ system 
using a full set of experimentally attainable parameters as discussed in Appendix D.  The simulation considered transmon and cavity frequencies in the standard 4-8 GHz range, Rabi rates $\Omega$ of 0.1-0.5 MHz %pump amplitude of $\xi_2<\xi_1=0.3$ (within the regime free of spurious heating effects~\cite{gao_programmable_2018}, adjusted for circuit parameters), 
(achieved with microwave drive amplitudes $\xi_2<\xi_1=0.3$ comparable to Ref.~\cite{gao_programmable_2018}), conservative internal $T_1$ times of 50, 25 and 800 $\mu$s for $\ket{e}_A$, $\ket{f}_A$ and $\ket{1}_B$~\cite{wang_schrodinger_2016} in addition to reservoir-induced Purcell effect, and a spurious %$\ket{g}\rightarrow\ket{e}$ transition rate of $\kappa/100$ for $R$ 
excited-state thermal population of 1\% in $R$ (comparable to Refs.~\cite{touzard_coherent_2018, wang_schrodinger_2016}) that dominates dephasing in $A$ and $B$.  The results for transferring a logical equator state (\textit{i.e.~}$(\ket{0}_L+\ket{1}_L)/\sqrt{2}$) are presented in Fig.~3(c).  For a wide range of $\chi_b$ and $\kappa$, the state transfer reaches within a few $\mu$s a fidelity of 89\%-93\% averaged over the six cardinal points of the Bloch sphere.  %Within the 6\% infidelity, about 0.5-2.5\% is intrinsic error shown in Fig.~3(c) inset, and the rest is due to decoherence.  
Leakage error out of the 12-dimension Hilbert space is not included, but its leading contribution from spurious transition of $R$ to its second excited state is estimated to be less than 0.2\%.  
Further improvement beyond these numerical results is possible if Purcell filters~\cite{houck_controlling_2008}, advanced thermalization techniques~\cite{yeh_microwave_2017}, or active/passive methods to cancel $\chi_b$~\cite{rosenblum_fault-tolerant_2018, zhang_suppression_2017} are employed.

\section{Implementation with bilinear interaction}

While our prescription for AQST explicitly requires three-body type of Hamiltonian interaction (Eq.~(4)), many physical systems only naturally support two-body (bilinear) interactions, such as Ising, two-mode squeezing, and Jaynes-Cummings type of interactions.  Nevertheless, it is possible to build composite degrees of freedom from multiple particles so that effective three-particle interactions can be achieved.  In the following, we provide a proof-of-principle example for AQST with only bilinear interaction.  %This is not an elaborated experimental proposal as the last section, but 

\begin{figure}[tbp]
    \centering
    \includegraphics[scale=1.05]{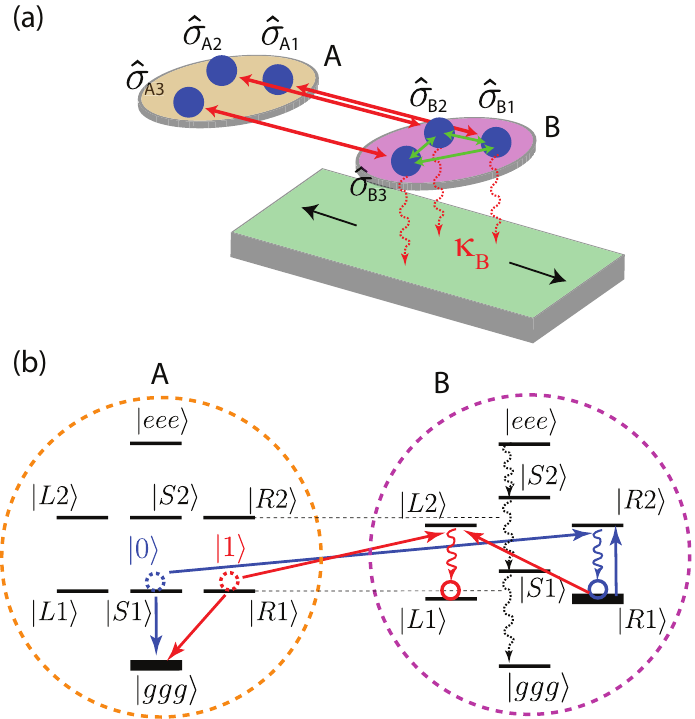}%[width=8.0cm]
    \caption{Autonomous transfer of a quantum state encoded in three atoms. (a) A schematic system with pair-wise interactions between atoms in $B$ (green arrows), swapping interaction between $\hat{\sigma}_{Ai}$ and $\hat{\sigma}_{Bi}$ ($i$ = 1, 2, 3) (red arrows), and collective dissipation of $B$ or $\hat{\mathcal{L}}=\hat{\sigma}^-_{B1}+\hat{\sigma}^-_{B2}+\hat{\sigma}^-_{B3}$ (red twisted lines).  (b) Energy level diagrams of $A$ and $B$.  The red and blue circles indicate the levels used for logical $\ket{0}$ and $\ket{1}$ when encoded in $A$ (dashed) or in $B$ (solid).  The thick black levels denote the vacuum states.  The red and blue arrows indicate the transfer paths for $\ket{0}$ and $\ket{1}$.
    }
    \label{bilinearAQST}
\end{figure}

We consider a system as in Fig.~\ref{bilinearAQST}, where the information emitter $A$ and receiver $B$ are each composed of three identical two-level atoms described by Pauli operators, ${\bf\hat\sigma}_{n1}$, ${\bf\hat\sigma}_{n2}$, ${\bf\hat\sigma}_{n3}$ ($n$ = $A$ or $B$).  The atomic states are $\ket{g}$ and $\ket{e}$ with transition energy $\hbar\omega$.
We consider a system Hamiltonian with swapping interactions between certain pairs of atoms
\begin{align}
\hat{H}=&\hbar\omega\sum_i \hat\sigma^z_{Ai} + \hbar\omega\sum_i \hat\sigma^z_{Bi} + \hbar J\sum_{i<j}\big(\hat\sigma^+_{Bi}\hat\sigma^-_{Bj}+\hat\sigma^-_{Bi}\hat\sigma^+_{Bj}\big) \nonumber\\
 &+\hbar g \sum_{i}\big(\sigma_{Ai}^-\sigma_{Bi}^+ +\sigma_{Ai}^+\sigma_{Bi}^- \big)
	\label{eq:AtomHamiltonian}
\end{align}
where $g\ll J\ll \omega$.  The swapping Hamiltonian is equivalent to the XY spin model and can arise, for example, from resonant dipolar interactions in Rydberg atoms~\cite{barredo_coherent_2015} or laser-driven interactions in trapped ions~\cite{jurcevic_quasiparticle_2014}.  
The three atoms in $B$ are subject to collective decay by emitting into the same reservoir with jump operator
\begin{equation}
\hat{L}_B=\sqrt{\kappa}\big(\sigma^-_{B1}+\sigma^-_{B2}+\sigma^-_{B3}\big)
\end{equation}  
%We will also include a similar collective decay for $A$ (\textit{i.e.}~$\hat{\mathcal{L}}_A=\sigma^-_{B1}+\sigma^-_{B2}+\sigma^-_{B3}$) for scalability considerations, which is not essential for the following transfer scheme from $A$ to $B$.

We define logical and vaccum states as (Fig.~\ref{bilinearAQST}(b)): %$\ket{0}_A=\ket{S1}_A$, $\ket{1}_A=\ket{R1}_A$, $\ket{\textrm{vac}}_A=\ket{ggg}_A$, $\ket{0}_B=\ket{R1}_B$, $\ket{1}_B=\ket{L1}_B$, $\ket{\textrm{vac}}_B=\ket{R1}_B$.
\begin{align}
\ket{0}_A=\ket{S1}_A, \,\,\ket{1}_A=\ket{R1}_A, \,\, \ket{\textrm{vac}}_A=\ket{ggg}_A \nonumber\\
\ket{0}_B=\ket{R1}_B, \,\, \ket{1}_B=\ket{L1}_B, \,\,\ket{\textrm{vac}}_B=\ket{R1}_B
\end{align}
%Here $\ket{S1}_n$, $\ket{L1}_n$ and $\ket{R1}_n$ ($n=A$ or $B$) 
Here
\begin{align}
\ket{S1}&=\ket{egg}+\ket{geg}+\ket{gge}\nonumber\\
\ket{L1}&=\ket{egg}+e^{i\frac{2}{3}\pi}\ket{geg}+e^{-i\frac{2}{3}\pi}\ket{gge}\nonumber\\
\ket{R1}&=\ket{egg}+e^{-i\frac{2}{3}\pi}\ket{geg}+e^{i\frac{2}{3}\pi}\ket{gge}
	\label{eq:states}
\end{align}
are symmetric, ``left-handed" and ``right-handed" states in the one-excitation manifold.
%which applies both $A$ and $B$ (hence subscripts are omitted).  
Due to the symmetric collective decay in $B$, $\ket{S1}_B$ is unstable but $\ket{L1}_B$ and $\ket{R1}_B$ are stable.  The three states for the two-excitation manifold are similarly defined as $\ket{S2}$, $\ket{L2}$, $\ket{R2}$, \textit{e.g.} $\ket{L2}=\ket{gee}+e^{i\frac{2}{3}\pi}\ket{ege}+e^{-i\frac{2}{3}\pi}\ket{eeg}$.  Note that $\ket{L2}$ and $\ket{R2}$ are also subject to decay, \textit{i.e.}~$\hat{L}_B\ket{R2}=\ket{R1}$. 

Starting from an initial state $\ket{\psi}_A\ket{\textrm{vac}}_B%=(\alpha\ket{S1}_A+\beta\ket{R1}_A)\otimes\ket{R1}_B
$, the relatively weak $\hbar g$ term resonantly couples the two initial logical states $\ket{S1(R1)}_A\ket{R1}_B$ to $\ket{ggg}_A\ket{R2(L2)}_B$ respectively% (red and blue arrows in Fig.~4(b))
, which allows the transfer of one collective excitation from $A$ to $B$. % while the qubit is encoded in the ``chirality" of the state.
(On the other hand, coupling to $\ket{R2(L2)}_A\ket{ggg}_B$ is off-resonant and can be neglected in RWA when $g\ll J$.) 
Subsequently %the dissipation 
$\hat{L}_B$ leads to decay from $\ket{ggg}_A\ket{L2(R2)}_B$ to steady states $\ket{ggg}_A\ket{L1(R1)}_B$ without acquiring the which-state information, completing the directional transfer to $\ket{\text{vac}}_A\ket{\psi}_B$.  

Our logical qubit in this construction is encoded in the ``chirality" of the superposition coefficients while both the Hamiltonian and the dissipation have the symmetry that preserves the total chirality.  The $\hbar g \sum_{i}\big(\sigma_{Ai}^-\sigma_{Bi}^+ 
+\sigma_{Ai}^+\sigma_{Bi}^- \big)$ term in Eq.~(\ref{eq:AtomHamiltonian}) plays the role of the transfer interaction as Eq.~(4).  It exchanges the chirality degree of freedom between $A$ and $B$ while simultaneously adding an energy excitation to a dissipative reservoir (subject to $\hat{L}_B$ decay), effectively achieving a three-body interaction necessary for AQST.

There are a few variations of this protocol worth considering.  The interactions within $B$ can alternatively use an Ising type of coupling such as $\sigma_{i}^z \sigma_{j}^z$.  Qubit encoding for $A$ and $B$ can also be made identical: $\ket{0}=\ket{R1}$ and $\ket{1}=\ket{L1}$ if the last term in Eq.~(\ref{eq:AtomHamiltonian}) is rewritten as $\hbar g \big(\sigma_{A1}^-\sigma_{B1}^+ +e^{i\frac{2}{3}\pi}\sigma_{A2}^-\sigma_{B2}^+ + e^{-i\frac{2}{3}\pi}\sigma_{A3}^-\sigma_{B3}^+ + c.c.\big)$.  For possible scaling up of the scheme into a chain of subsystems, inter-atom couplings and collective dissipation can be introduced to $A$ to enable it as a receiver of quantum state from further upstream emitters.

\section{General conditions}
So far we have focused on one type of strategy for AQST by synthesizing a one-step quantum jump.  More generally, AQST can be realized through a more complex trajectory with many jumps and/or jump operators.  In this section, we discuss the conditions for autonomous transfer of one qubit of information.  %We consider a simple class of scenario: 

Consider a system consisting of $A$, $B$, and an auxiliary subsystem $C$ that has Hamiltonian $\hat{H}$ and interacts with $m$ independent Markovian reservoirs described by jump operators $\hat{L}_\mu$, $\mu\in[1,m]$. In order for a qubit to be transferred from $A$ and eventually stored in $B$, first of all, a two-dimensional stationary dark-state manifold $\mathcal{M}_B=\text{span}\big\{\ket{\phi_{0,B}}, \ket{\phi_{1,B}}\big\}$ is needed to encode information locally in $B$ with
%, where $\ket{\phi_{k,B}}=\ket{\text{vac}}_A\ket{k}_B\ket{x}_C$ ($k\in\{0,1\}$), 
\begin{align}
&\ket{\phi_{\sigma,B}}=\ket{\text{vac}}_A\ket{\sigma}_B\ket{\text{vac}}_C, \,\,  \forall \sigma\in\{0,1\}\nonumber\\
&\hat{L}_\mu\ket{\phi_{\sigma,B}}=0, \,\forall\mu \,\,\,\, \textrm{and} \,\,\,\, \hat{H}\ket{\phi_{\sigma,B}}=E_{\sigma,B}\ket{\phi_{\sigma,B}}
\end{align}
%We will use this dark state manifold as the encoding subspace $\mathcal{H}_B=\text{span}\big\{ |\phi_{k}\rangle\big\}$ that receives the quantum state.
Secondly, all basis states of the initial state manifold  $\mathcal{M}_A=\text{span}\big\{\ket{\phi_{0,A}},\ket{\phi_{1,A}}\big\}$, with $\ket{\phi_{\sigma,A}}=\ket{\sigma}_A\ket{\text{vac}}_B\ket{\text{vac}}_C$, should be attracted onto $\mathcal{M}_B$ at long times.  %The detailed dynamics can be described by an ensemble of quantum trajectories. %each including a series of quantum jumps from $\{\hat{L}_{\mu}\}$ intertwined with no-jump evolutions %for time interval $\tau$: $\hat{L}_{0}(\tau)=\exp{\big(-i\hat{H}/\hbar-\sum_{1}^{m}\frac{1}{2}\hat{L}_\mu^\dagger\hat{L}_\mu\big)\tau}$ %with $\hat{H}_{\text{eff}}=\hat{H}-i\hbar\sum_\mu\frac{1}{2}\hat{L}_\mu^\dagger\hat{L}_\mu$ over time step $\tau$.  
The third requirement is that orthogonal states in $\mathcal{M}_A$ remain orthogonal throughout any possible quantum trajectories evolving towards $\mathcal{M}_B$.  This is necessary and sufficient to ensure no leakage of quantum information to the environment, which is equivalent to meeting the Knill-Laflamme quantum error correction criteria~\cite{knill_theory_1997} at all times%(See appendix for a more systematic definition of these requirements for a more general transfer of arbitrary dimension quantum state with arbitrary number of jump operators.)
\begin{equation}
\bra{\phi_{\sigma,A}}\hat{\mathcal{K}}_i^\dagger\hat{\mathcal{K}}_j\ket{\phi_{\sigma',A}}=\eta_{ij}\delta_{kl}, \,\,\, \forall \sigma,\sigma',i,j
	\label{eq:condition}
\end{equation}
where $\hat{\mathcal{K}}_i$ and $\hat{\mathcal{K}}_j$ are any possible Kraus operators after a given evolution time, and $\eta_{ij}$ is Hermitian.  

\begin{figure}[tbp]
    \centering
    \includegraphics[scale=0.95]{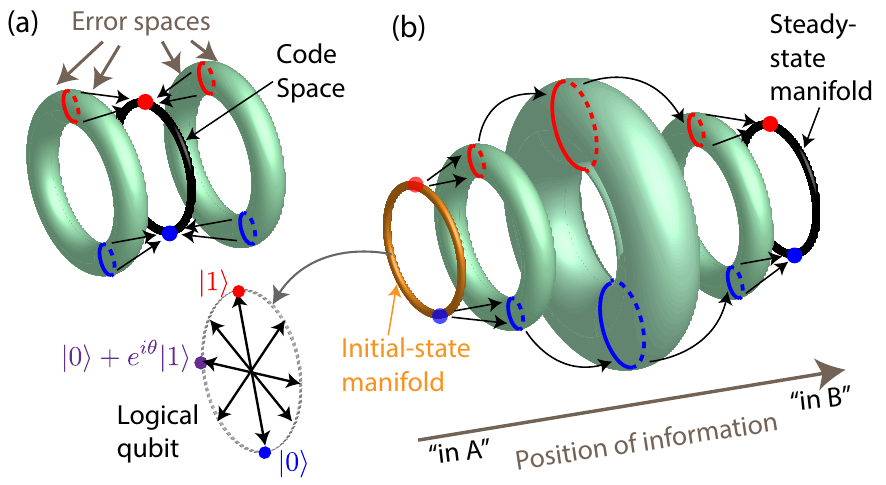}%[width=8.0cm]
    \caption{Conceptual comparison of (a) AQEC and (b) AQST.  In either case, the black ring represents a 2D steady-state manifold encoding a qubit (``a Bloch sphere seen from low dimension"), where the radial angle on the ring represents the qubit state.  Any translation or expansion/shrinkage of the ring create an alternative encoding space (error space) for the qubit.  Each thickened green ring represents a mixture of multiple error spaces.  A state depicted by a red or blue band is mixed among different error spaces but still contains a pure logical qubit state.  In (a) AQEC, dissipation maps a few neighboring error spaces (and their mixtures) back to the code space.  In (b) AQST, dissipation maps $\mathcal{M}_A$ and any mixture of intermediate code spaces to $\mathcal{M}_B$.
    }
    \label{AQEC-AQST}
\end{figure}

The AQST discussed in this Article is intrinsically connected to AQEC~\cite{ahn_continuous_2002,lihm_implementation-independent_2018,kerckhoff_designing_2010,kapit_hardware-efficient_2016,reiter_dissipative_2017,cohen_autonomous_2017,albert_pair-cat_2019}, as reflected by the above requirements similar to (but stronger than) that of AQEC~\cite{lihm_implementation-independent_2018}: The initial manifold $\mathcal{M}_A$ can be viewed as an error space that is being continuously mapped back to the correct code space $\mathcal{M}_B$ through dissipation engineering in AQEC.  The difference is technical but yet distinct: AQEC is designed to recover information from an adjacent error space that is typically separated from the code space by the perturbation of a single natural error.  On the other hand, AQST seeks to transport information from an initial space as distant from the final code space as necessary to store the logical qubit in a different physical subsystem (Fig.~\ref{AQEC-AQST}).
As a result, the Kraus operators in general involve a series of quantum jumps from $\{\hat{L}_{\mu}\}$ intertwined with no-jump evolutions $\hat{L}_{0}(\tau)=\exp{\big(-i\hat{H}/\hbar-\sum_{1}^{m}\frac{1}{2}\hat{L}_\mu^\dagger\hat{L}_\mu\big)\tau}$, making Eq.~(\ref{eq:condition}) fairly difficult to use in practice. A helpful strategy to design AQST schemes is to conceptually divide the global system into two independent degrees of freedom $L$ and $P$, a ``logical'' ($L$) qubit mode that contains the information $\ket{\psi}$ and is associated with certain symmetry, and a ``position'' ($P$) mode that marks where the information is.
%The initial states can be relabeled as:
%\begin{align}
%|0\rangle_A|\text{vac}\rangle_B \equiv |0\rangle_L|A\rangle_P, \,\,\,\,
%|1\rangle_A|\text{vac}\rangle_B \equiv |1\rangle_L|A\rangle_P\nonumber\\
%|\text{vac}\rangle_A|0\rangle_B \equiv |0\rangle_L|B\rangle_P, \,\,\,\,
%|\text{vac}\rangle_A|1\rangle_B \equiv |1\rangle_L|B\rangle_P
%\end{align}
%We then design the global accessible Hilbert space to be divisible as $L\otimes P$, and attempt 
We then engineer $\hat{H}$ and $\hat{L}_\mu$ to respect sufficient symmetry and drive non-reciprocal interactions in mode $P$ only, and therefore maintain the density matrix of the global system in a separable form of
\begin{equation}
\hat{\rho}=\bigg(|\psi\rangle_L\langle\psi|_L\bigg)\otimes 
\bigg(\sum_i p_i|i\rangle_P\langle i|_P\bigg)
\end{equation}
Here $|i\rangle_P$ are eigenstates of the position mode including (but not limited to) $|A\rangle_P$ and $|B\rangle_P$.
%This can be achieved if the jump operators $\hat{L_\mu}$ and Hamiltonian $\hat{H}$ respects sufficient symmetry to drives dynamics in mode $P$ only, i.e.~$\big[\hat{o},|\psi\rangle_L\langle\psi|_L\big]=0$ and $\big[\hat{H},|\psi\rangle_L\langle\psi|_L\big]=0$. 

\section{Outlook}
We have shown that it is possible to construct a dissipative quantum channel where logical qubit states are autonomously fed forward from one subsystem to the next.  This can be achieved in example protocols by synthesizing a one-step relaxation process from a pair of intermediate states to the final states, and more generally by engineering dissipation that transfers excitations while maintaining orthogonality of underlying logical states.  AQST does not entangle the source qubit and the receiving qubit, but entanglement with any external party is preserved.  AQST can be implemented in a variety of local or remote physical settings, and is achievable in circuit QED under current experimental capabilities.   

Looking forward, an intrinsically directional but still information-preserving channel may be used to enforce hierarchy and improve isolation in modular architectures of quantum computation~\cite{jiang_distributed_2007, monroe_large-scale_2014}.  The ability to implement essential QIP operations without time-dependent external control also leads to potential savings in arbitrary-waveform control electronics, thus addressing one of the scalibility bottlenecks for quantum computing architectures~\cite{bardin_2019, mcdermott_quantumclassical_2018}.  It will also be interesting to explore AQST schemes to include protection against errors, for example, in multi-cavity bosonic states~\cite{wang_schrodinger_2016, albert_pair-cat_2019}. Beyond gate-based QIP, the use of dissipative engineering for state transfer may also be integrated into dissipative quantum computation~\cite{verstraete_quantum_2009}.  Finally, AQST naturally implements irreversible a classical OR gate (\textit{e.g.}~let $\ket{\text{vac}}_A\equiv\ket{2}_A$ and $\ket{\text{vac}}_B\equiv\ket{0}_B$ in Eq.~(\ref{eq:3x2jump}), the output of $B$ equals $A$ OR $B$)
, which may inspire ways to combine quantum and classical logic in the same system.

%\subsection*{}
\textit{note added--} During the revision of the manuscript, the authors became aware of two related work~\cite{li_directional_2019, matsuzaki_one-way_2018}, which propose different dissipative protocols to implement directional quantum state transfer in specific systems.  In contrast, this article more generally addresses the minimum resources, the generic strategy, the applicability in various physical settings for directional quantum state transfer, and also builds a detailed experimental plan towards realizing AQST in circuit QED.  

\begin{acknowledgments}
We thank Aashish Clerk, Mazyar Mirrahimi, Liang Jiang, Xiaowei Deng and Serge Rosenblum for helpful discussions.  This research was supported by U.S. Army Research Office (W911NF-17-1-0469) 
and Air Force Office of Scientific Research (FA9550-18-1-0092). 
\end{acknowledgments}

\begin{appendix}
\section{Instrinsic Infidelity of AQST via a Traveling Mode}
Following Equations (5)-(10), before any quantum jump happens, the dynamics of the global system can be described by a stochastic wave function evolving according to the Schrodinger equation
\begin{equation}
    i\hbar\frac{d\ket{\phi(t)}}{dt}=\hat{H}_\text{eff}\ket\phi
    \label{eq:Schrodinger}
\end{equation}
with non-Hermitian effective Hamiltonian~\cite{carmichael_quantum_1993, pinotsi_single_2008}:
\begin{align}
     \hat{H}_\text{eff}=&\sum_{\sigma=0}^1 \hbar\lambda\big(\hat{a}_\sigma^\dagger\hat{A}_\sigma+\hat{a}_\sigma\hat{A}^\dagger_\sigma)
     -i\hbar\sqrt{\kappa_a\kappa_b}\hat{a}_\sigma\hat{b}_\sigma^\dagger-i\hbar\frac{\kappa_a}{2} \hat{a}_\sigma^\dagger\hat{a}_\sigma \nonumber\\
     &-i\hbar\frac{\kappa_b}{2} \hat{b}_\sigma^\dagger\hat{b}_\sigma 
     +\hbar\Omega\big(\hat{b}_\sigma^\dagger\hat{B}_\sigma\hat{r}+\hat{b}_\sigma\hat{B}^\dagger_\sigma\hat{r}^\dagger) -i\hbar\frac{\gamma}{2}\hat{r}^\dagger\hat{r}
     \label{eq:H_eff}
\end{align}
where we introduced notations $\hat{A}_\sigma=\ket{\text{vac}}_A\bra{\sigma}_A$,  $\hat{B}_\sigma=\ket{\text{vac}}_B\bra{\sigma}_B$ ($\sigma=0,1$), and $\hat{r}=\ket{g}_R\bra{e}_R$. 

We consider the parameter regime of $\lambda\ll\kappa_a,\kappa_b,\gamma$ to satisfy the adiabaticity condition, and enforce $\kappa_b=4\Omega^2/\gamma$ to satisfy the impedance matching condition.  Starting from an initial wave function of  $\ket{\phi(t=0)}=\ket{\psi}_A\ket{\text{vac}}_a\ket{\text{vac}}_b\ket{\text{vac}}_B\ket{g}_R$, $\ket{\psi}=\alpha\ket{0}+\beta\ket{1}$, since the dynamics associated with transferring logical $\ket{0}$ and $\ket{1}$ are exactly mirrored and do not interact with each other, we have:
\begin{align}
    \ket{\phi(t)}=&C_0(t)\ket{\psi}_A\ket{\text{vac}}_a\ket{\text{vac}}_b\ket{\text{vac}}_B\ket{g}_R\nonumber\\
    &+C_1(t)\ket{\text{vac}}_A\ket{\psi}_a\ket{\text{vac}}_b\ket{\text{vac}}_B\ket{g}_R\nonumber\\
    &+C_2(t)\ket{\text{vac}}_A\ket{\text{vac}}_a\ket{\psi}_b\ket{\text{vac}}_B\ket{g}_R\nonumber\\
    &+C_3(t)\ket{\text{vac}}_A\ket{\text{vac}}_a\ket{\text{vac}}_b\ket{\psi}_B\ket{e}_R
    \label{eq:wavefunction}
\end{align}
There exists a steady state solution for the above wave function, $\ket{\phi_\infty}$, with:
\begin{align}
    &C_1=-\frac{2i\lambda}{\kappa_a}, \,\,\, C_2=\frac{2i\lambda}{\sqrt{\kappa_a\kappa_b}}, \,\,\, C_3=\frac{2\lambda}{\sqrt{\kappa_a\gamma}}, \,\,\,  \nonumber\\
    &C_0=\sqrt{1-4\bigg(\frac{\lambda^2}{\kappa_a^2}+\frac{\lambda^2}{\kappa_a\kappa_b}+\frac{\lambda^2}{\kappa_a\gamma}\bigg)}\approx1
\end{align}
It is easy to verify that $\hat{H}_\text{eff}\ket{\phi_\infty}\approx 0$
%\begin{equation}
%   \hat{H}_\text{eff}\ket{\phi_\infty}=-\frac{2i\lambda^2}{\kappa_a}\ket{\psi}_A\ket{\text{vac}}_a\ket{\text{vac}}_b\ket{\text{vac}}_B\ket{g}_R\approx0
%\end{equation}
to first order in $\lambda$.  It can be further checked that, up to second order in $\lambda$, $d\ket{\phi_\infty}/dt=0$ after renormalization. $\ket{\phi_\infty}$ is the quasi-steady state that the system asymptotically approaches before any quantum jump occurs. Impedance msatching is realized because $\ket{\phi_\infty}$ is a dark state for jump operators $\hat{L}_0$ and $\hat{L}_1$:  $\hat{L}_\sigma\ket{\phi_\infty}= (\sqrt{\kappa_a}\hat{a}_\sigma+\sqrt{\kappa_b}\hat{b}_\sigma)\ket{\phi_\infty}=0$.  Therefore once the wave function evolves past its initial transient dynamics, information loss due to the reflection at cavity $b$ can no longer occur.  Adiabaticity is realized with the excitation probability in cavities $a$ and $b$ kept small at all times ($|C_1|^2, |C_2|^2\ll 1$), suppressing the occurrence probability of $\hat{L}_0$ and $\hat{L}_1$ during the transient evolution towards $\ket{\phi_\infty}$.

The intrinsic infidelity of this AQST protocol can be evaluated by solving the (transient) time-dependent coefficients in Eq.~(\ref{eq:wavefunction}).  Since the asympototic state has $|C_1|^2$, $|C_2|^2$, $|C_3|^2\ll 1$ up to first order in $\lambda$, we assume $C_0(t)=1-o(\lambda^2/\kappa_a^2)-o(\lambda^2/\kappa_b^2)-o(\lambda^2/\gamma^2)\approx 1$.  From Eqs.~(\ref{eq:Schrodinger}), (\ref{eq:H_eff}) and (\ref{eq:wavefunction}), we arrive at equations of motion:
\begin{align}
    \dot{C_1}&=-i\lambda-\frac{\kappa_a}{2} C_1\\
    \dot{C_2}&=-\sqrt{\kappa_a\kappa_b} C_1 -\frac{\kappa_b}{2}C_2 -i\frac{\sqrt{\kappa_b\gamma}}{2} C_3\\
    \dot{C_3}&=-i\frac{\sqrt{\kappa_b\gamma}}{2} C_2 -\frac{\gamma}{2} C_3
\end{align}
where we have used $\Omega=\sqrt{\kappa_b\gamma}/2$. The solution has the following form to be consistent with the steady-state solution at $t\to\infty$:
\begin{align}
    C_1(t)&=-\frac{2i\lambda}{\kappa_a} \big(1-e^{-\frac{\kappa_a}{2} t}\big)\\
    C_2(t)&=\frac{2i\lambda}{\sqrt{\kappa_a\kappa_b}} \big(1-x_2 e^{-\frac{\kappa_a}{2} t}-y_2 e^{-\kappa' t}-z_2 e^{-\gamma' t} \big)\\
    C_3(t)&=\frac{2\lambda}{\sqrt{\kappa_a\gamma}} \big(1-x_3 e^{-\frac{\kappa_a}{2} t}-y_3 e^{-\kappa' t}-z_3 e^{-\gamma' t} \big)
\end{align}
where
\begin{equation}
    x_2+y_2+z_2=x_3+y_3+z_3=1
\end{equation}
to satisfy the initial condition at $t=0$. 
Plug Eqs.~(A8)-(A10) into Eqs.~(A6)-(A7), and collect coefficients for the three different exponential components, the solutions are:
%\begin{align}
%    \kappa_a x_2 &= \kappa_b (x_2+x_3-2)\\
%    2\kappa' y_2 &= \kappa_b (y_2+y_3)\\
%    2\gamma' z_2 &= \kappa_b (z_2+z_3)\\
%    \kappa_a x_3 &= \gamma (x_3-x_2)\\
%    2\kappa' y_3 &= \gamma (y_3-y_2)\\
%    2\gamma' z_3 &= \gamma (z_3-z_2)
%\end{align}
%Solving Eq.~(S12)-(S17), we get:
\begin{align}
    \kappa',\gamma' &= \frac{1}{4}\bigg[(\kappa_b+\gamma)\pm\sqrt{(\kappa_b+\gamma)^2-8\gamma\kappa_b}\bigg]
    \label{eq:timescale}\\
    x_2 &= \frac{2\kappa_b(\gamma-\kappa_a)}{2\kappa_b\gamma+\kappa_a^2-\kappa_a\kappa_b-\kappa_a\gamma}\\
    x_3 &= \frac{2\kappa_b\gamma}{2\kappa_b\gamma+\kappa_a^2-\kappa_a\kappa_b-\kappa_a\gamma}\\
    y_2, z_2 &= \frac{1}{2}\bigg[(1-x_2)\pm\frac{(\kappa_b+\gamma)+(2\kappa_a-\kappa_b-\gamma)x_2}{\sqrt{(\kappa_b+\gamma)^2-8\kappa_b\gamma}}\bigg]\\
    y_3, z_3 &= \frac{1}{2}\bigg[(1-x_3)\pm\frac{(\kappa_b+\gamma)+(2\kappa_a-\kappa_b-\gamma)x_3}{\sqrt{(\kappa_b+\gamma)^2-8\kappa_b\gamma}}\bigg]
\end{align}

From Eq.~(\ref{eq:timescale}), we note that when $\gamma/\kappa_b < 2\sqrt2-1\approx1.83$ (or equivalently when $\gamma/\Omega < 2\sqrt{2\sqrt{2}-1}\approx2.71$), the dynamics between %$\ket{\psi}_b\ket{\text{vac}}_B\ket{g}_R$ and $\ket{\text{vac}}_b\ket{\psi}_B\ket{e}_R$
$C_2$ and $C_3$ corresponds to under-damped oscillation with exponential decay rate of $\Re(\kappa')=\Re(\gamma')=(\kappa_b+\gamma)/4$.  When $\gamma/\kappa_b > 2\sqrt2-1$, the oscillation is over-damped, and system approaches the quasi-steady state under a linear combination of three exponential time scales, $\kappa_a$, $\kappa'$ and $\gamma'$.  In the limit of $\gamma\gg\kappa_b$,
\begin{align}
    \kappa' &\approx \kappa_b \bigg(1-\frac{\kappa_b}{\gamma}\bigg) \approx \kappa_b\\
    \gamma' &\approx \frac{\gamma}{2} \bigg(1-\frac{\kappa_b}{\gamma}\bigg) \approx \frac{\gamma}{2}
\end{align}
and $x_2 \approx \frac{2\kappa_b}{2\kappa_b-\kappa_a}$, $y_2 \approx \frac{\kappa_a}{2\kappa_b-\kappa_a}$, $z_2 \approx 0$. This is the limit when the dynamics of the reservoir (\textit{i.e.~}the $C_3$ term) can be adiabatically eliminated, and the state of the receiving cavity $b$ is only governed by two time scales (the $C_2$ term): $\kappa_a/2$ and $\kappa_b$.

The total probability rate for quantum jumps $\hat{L}_0$ and $\hat{L}_1$ to occur at any given time is 
\begin{align}
    \sum_{\sigma=0}^1\bra{\phi(t)}\hat{L}_\sigma^\dagger\hat{L}_\sigma\ket{\phi(t)}=|\sqrt{\kappa_a} C_1(t)+\sqrt{\kappa_b} C_2(t)|^2 \nonumber\\
    =\bigg|\frac{2i\lambda}{\sqrt{\kappa_a}}
    \big[(1-x_2)e^{-\frac{\kappa_a}{2}t}-y_2 e^{-\kappa' t}-z_2 e^{-\gamma' t} \big]\bigg|^2
\end{align}
The infidelity $1-\mathcal{F}$ due to information leakage during the transient stage is given by a time-integral of the above probability rate:
\begin{align}
    1-\mathcal{F} &= \int_{t=0}^{\infty} \sum_{\sigma=0}^1\bra{\phi(t)}\hat{L}_\sigma^\dagger\hat{L}_\sigma\ket{\phi(t)} \mathrm{d}t
\nonumber\\
    &=\frac{4\lambda^2}{\kappa_a} \int_{t=0}^{\infty}\big|
    (1-x_2)e^{-\frac{\kappa_a}{2}t}-y_2 e^{-\kappa' t}-z_2 e^{-\gamma' t}\big|^2 \mathrm{d}t
    \label{eq:infidelity}
\end{align}
In the limit of $\gamma\gg\Omega$ so that $\lambda\ll\kappa_a, \kappa_b \ll \gamma$:
\begin{align}
    1-\mathcal{F} &=\frac{4\lambda^2}{\kappa_a} \int_{t=0}^{\infty}\bigg|
    \frac{\kappa_a}{2\kappa_b-\kappa_a}(-e^{-\frac{\kappa_a}{2}t}+e^{-\kappa_b t})\bigg|^2 \mathrm{d}t \nonumber\\
%    &=\frac{4\lambda^2 \kappa_a}{(2\kappa_b-\kappa_a)^2}\big[\frac{1}{\kappa_a}+\frac{1}{2\kappa_b}-\frac{2}{\kappa_a/2+\kappa_b}\big]
%\\&= \frac{4\lambda^2}{(2\kappa_b-\kappa_a)^2} \frac{\kappa_b (\kappa_a+2\kappa_b)+\kappa_a(\kappa_a/2+\kappa_b)-4\kappa_a\kappa_b  }{\kappa_b(\kappa_a+2\kappa_b)}
%\\&= \frac{4\lambda^2}{(2\kappa_b-\kappa_a)^2} \frac{\kappa_b (2\kappa_b)+\kappa_a(\kappa_a/2)-2\kappa_a\kappa_b}{\kappa_b(\kappa_a+2\kappa_b
    &=\frac{2\lambda^2}{\kappa_b(\kappa_a+2\kappa_b)}
    \label{eq:infidelity2}
\end{align}

The AQST fidelity shown in Fig.~2(c) is the result of numerically solving the Lindblad master equation and is not subject to the assumption of $\lambda\ll\kappa_a,\kappa_b,\gamma$ as is the analytic solutions described in this section [Eq.~(\ref{eq:infidelity}) and Eq.~(28-32)].  As expected, the two agrees in the small $\lambda$ regime.  Somewhat surprisingly, as shown in Fig.~2(c), for given $\lambda$, $\kappa_a$, $\kappa_b$, choosing $\gamma\gg\kappa_b$ is not only unnecessary but also not optimal for state transfer fidelity. Instead, highest fidelity is achieved near critical damping: $\gamma/\kappa_b = 2\sqrt2-1$.  This can be verified by evaluating Eq.~(\ref{eq:infidelity}) for different $\gamma$.  We also remark that our numerical results in Fig.~2(c) treats the specific case of $\kappa_a=\kappa_b$, which is an arbitrary choice and by no means optimal.  In fact, in practice, the regime of $\lambda<\kappa_a<\kappa_b$ may be beneficial as the relative gain in transfer speed ($4\lambda^2/\kappa_a$) by reducing $\kappa_a$ outweighs the relative increase in infidelity (\textit{i.e.} Eq.~(\ref{eq:infidelity2})).

%\section{Circuit QED implementation of AQST}
\section{Driven Josephson Circuit Hamiltonian}
The cQED Hamiltonian for the circuit in Fig.~3(a) incorporating two microwave drives with angular frequencies $\omega_1$, $\omega_2$ and amplitudes $\epsilon_1$, $\epsilon_2$ can be written as~\cite{leghtas_confining_2015}
\begin{align}
&\hat{H}=\hbar\omega_A \hat{a}^\dagger \hat{a}+\hbar\omega_B \hat{b}^\dagger \hat{b}+\hbar\omega_R \hat{r}^\dagger \hat{r}\nonumber\\
&-\sum_{i=\text{I, II}}E_{Ji}\big[\cos\big(\hat{\varphi}_i\big) + \frac{\hat{\varphi}_i^2}{2}\big]
+ \sum_{k=1,2}\hbar\epsilon_k \cos(2\omega_k t)(\hat{r} + \hat{r}^\dagger \big)
\label{eq:cQED_Hamiltonian}
\end{align}
where $\hat{a}^\dagger$, $\hat{b}^\dagger$ and $\hat{r}^\dagger$ are creation operators of LC oscillator modes that are closely associated with $A$, $B$ and $R$. $E_{Ji}$ is the Josephson inductance of junction $i$ (= I or II).  The phases across the junctions I and II are given by
\begin{align}
\hat{\varphi}_i = \Phi_{Ai}\big(\hat{a}^\dagger+\hat{a}\big)
+\Phi_{Bi}\big(\hat{b}^\dagger+\hat{b}\big)
+\Phi_{Ri}\big(\hat{r}^\dagger+\hat{r}\big)
\end{align}
$\Phi_{Xi}$ is the zero-point flux fluctuation of mode $X$ (=$A$, $B$, or $R$) across junction $i$.  We use Junction I as a resource of three-body nonlinear coupling with relatively large $\Phi_{A\text{I}}\Phi_{B\text{I}}\Phi_{R\text{I}}$ product.  Junction II is mostly for the purpose of providing anharmonicity to make $A$ a usable qutrit, whose coupling to $B$ is negligible (\textit{i.e.~}$\Phi_{B\text{I}}\approx 0$) and contribute relatively little to the state transfer process.  Because the junctions I and II are located within transmons $R$ and $A$ respectively, $\Phi_{R\text{I}}, \Phi_{A\text{II}} \gg$ all other $\Phi_{Xi}$'s.
%\begin{align}
%\hat{H}=&\hbar\omega_A \hat{a}^\dagger \hat{a}+\hbar\omega_B \hat{b}^\dagger \hat{b}+\hbar\omega_R \hat{r}^\dagger \hat{r}\nonumber\\
%&-\sum_{i=\text{I},\text{II}}\frac{E_{Ji}}{24}\big[\Phi_{Ai}\big(\hat{a}^\dagger+\hat{a}\big)
%+\Phi_{Bi}\big(\hat{b}^\dagger+\hat{b}\big)
%+\Phi_{Ri}\big(\hat{r}^\dagger+\hat{r}\nonumber\\
%&+\xi_1 e^{-i\omega_1 t}+\xi_1^* e^{i\omega_1 t}
%+\xi_2 e^{-i\omega_2 t}+\xi_2^* e^{i\omega_2 t}\big)\big]^4
%\end{align}

%The last term in Eq.~(\ref{eq:cQED_Hamiltonian}) describes two microwave drives with frequencies (This has to be rearranged... It's difficult to introduce chi's and alphas here, I don't think you need to write out $\omega$ here)
%\begin{gather}
%\omega_1 = \omega_B + \omega_R - \omega_A + \delta_1\\
%\omega_2 = 2 \omega_A - \chi_A - \omega_R + \delta_2
%\end{gather}

Using the same procedure as performed in the supplementary of \cite{leghtas_confining_2015} we can move the drive terms in Eq.~(\ref{eq:cQED_Hamiltonian}) into the cosine expansion (arriving at Eq.~(\ref{eq:4thorder}) if the readers wish to skip this part of derivation).  First we move to an effective Hamiltonian by introducing a non-Hermitian loss term for our reservoir mode, and Taylor expanding $\cos\hat{\varphi}_i$ to the 4th order:
\begin{align}
\hat{H}&=\hbar\omega_A \hat{a}^\dagger \hat{a}+\hbar\omega_B \hat{b}^\dagger \hat{b}
+\hbar(\omega_R+ i \kappa) \hat{r}^\dagger \hat{r}
\nonumber\\
&-\sum_{i=I, II}\frac{E_{Ji}}{24}\big[\Phi_{Ai}\hat{a}
+\Phi_{Bi}\hat{b}
+\Phi_{Ri}\hat{r}+ h.c.\big]^4 \nonumber\\
&+ \sum_{k=1,2}\hbar\epsilon_k \cos(2\omega_k t)(\hat{r} + \hat{r}^\dagger \big)
%\label{eq:cQED_Hamiltonian2}
\end{align}
Now we apply displacement transformation:
\begin{equation}
    \hat{r}\rightarrow\hat{r}+\xi_1 e^{-i\omega_1 t}+\xi_2 e^{-i\omega_2 t}
\end{equation}
using two independent unitary operators with complex displacement amplitudes $\xi_1$ and $\xi_2$ to be determined later: %$\hat{H}_\text{eff}\rightarrow \hat{U_1}^\dagger\hat{U_2}^\dagger \hat{H}_\text{eff}\hat{U_1}^\dagger\hat{U_1}$ with the frequencies of the two drive tones but arbitrary amplitudes:
\begin{align}
\hat{U}_{1,2} = \hat{D}(-\xi_{1,2} e^{-i \omega_{1,2} t})
\end{align}
where the displacement operator is performed on the reservoir mode
\begin{equation}
    \hat{D}(\alpha)=e^{\alpha\hat{r}^\dagger-\alpha^*\hat{r}}
\end{equation}
The Hamiltonian Eq.~(\ref{eq:cQED_Hamiltonian}) is now:
\begin{widetext}
\begin{align}\label{eq:first_transform}
%\begin{split}
    \hat{H} &= \hbar\omega_A \hat{a}^\dagger \hat{a}+\hbar\omega_B \hat{b}^\dagger \hat{b} 
    -\sum_{i=I, II}\frac{E_{Ji}}{24}\big[\Phi_{Ai}\hat{a}
    +\Phi_{Bi}\hat{b}
    +\Phi_{Ri}\big(\hat{r}+\xi_1 e^{-i\omega_1 t}+\xi_2 e^{-i\omega_2 t} \big)+ h.c.\big]^4\nonumber\\
    &+\sum_{k=1,2} \bigg\{\hbar(\omega_R + i\kappa)(\hat{r}^\dagger + \xi_k^* e^{i\omega_k t})(\hat{r} + \xi_k e^{-i \omega_k t}) \nonumber\\
    &+\hbar\epsilon_k 2 Re\{\epsilon e^{-i\omega_k t}\}(\hat{r} + \xi^* e^{i\omega_k t} + \hat{r}^\dagger + \xi e^{-i \omega_k t} \big) + i \hbar [(-\dot{\xi}_k + i\omega_k \xi_k) e^{-i \omega_k t}\hat{r}^\dagger  + (\dot{\xi}_k^* + i\omega_k \xi_k^*) e^{i \omega_k t}\hat{r}]\bigg\}
%    \label{eq:first_transform}
%\end{split}
\end{align}
\end{widetext}
We can rewrite this, collecting terms inside $\{...\}$ into the form of %that contain $\hat{r}^\dagger$,  terms that contain $\hat{r}$, terms that contain $\hat{r}^\dagger \hat{r}$ and constant terms.
\begin{align}
%    \hat{H} =  \hbar\omega_A \hat{a}^\dagger \hat{a}+\hbar\omega_B \hat{b}^\dagger \hat{b}
    \lambda^* \hat{r}^\dagger
    + \lambda \hat{r} + (\omega_R + \delta \omega_R + i\kappa) \hat{r}^\dagger \hat{r} + constants 
    \label{eq:reorg}
\end{align}
Here, $\delta \omega_R$ is a shift in the bare frequency of the mode, and $\lambda$ can be written as:
\begin{align}
    \lambda =& \sum_{k=1,2}2 Re\{\epsilon_k e^{-i\omega_k t}\} + i \hbar (-\dot{\xi_k} + i\omega_k \xi_k) e^{-i \omega_k t}\nonumber\\
    &+ \hbar(\omega_r + i\kappa)(\hat{r}^\dagger + \xi_k^* e^{i\omega_k t})
\end{align}
We want to choose $\xi_{1,2}$ and $\omega_{1,2}$ such that $\lambda = \lambda^* = 0$ these terms disappear. If we rewrite
\begin{align}
\tilde{\xi}_{1,2} \equiv \xi_{1,2} e^{-i \omega_{1,2} t}
\end{align}
This gives us two independent conditions for $\xi_{1,2}$.
\begin{align}
\dot{\tilde{\xi}}_{1,2} = (i (\omega_a - \omega_{1,2}) + \kappa_a) \tilde{\xi}_{1,2} - \frac{2i}{\hbar} Re\{\epsilon_{1,2} e^{-i\omega_{1,2} t}\}
\end{align}
On a time scale of $1/\kappa$ we can approximate this as
\begin{align}
\xi_{1,2} \approx i\epsilon_{1,2} / (\kappa_a + i(\omega_a - \omega_{1,2}))
\end{align}
These values are dependant on the detunings between the reservoir mode and the drive tones, the reservoir linewidth, and the strength of the drive tones. Keeping terms up to 4th order in the expansion of the junction energies our Hamiltonian now looks like:
\begin{align}
\hat{H}&=\hbar\omega_A \hat{a}^\dagger \hat{a}+\hbar\omega_B \hat{b}^\dagger \hat{b} +\hbar\omega_R \hat{r}^\dagger \hat{r} \nonumber\\
&-\sum_{i=I, II}\frac{E_{Ji}}{24}\big[\Phi_{Ai}\hat{a}
+\Phi_{Bi}\hat{b}
+\Phi_{Ri}\big(\hat{r}+\tilde{\xi}_1
+\tilde{\xi}_2 \big)+ h.c.\big]^4
\label{eq:4thorder}
\end{align}
For simplicity of the derivation, we assumed that the drive is equally effective in coupling to Junctions I and II.  This is not essential for the experiment as Junction II contributes very little to the conversion Hamiltonian in the state transfer. 

%Here we have taken the cosine expansion of Josephson energy to the 4$^{th}$ order, with the drive terms absorbed into the Josephson nonlinear terms after a displacement transformation (see \textit{e.g.} supplementary info of Ref.~\cite{leghtas_confining_2015,gao_programmable_2018}). 
The 4th order expansion in Eq.~(\ref{eq:4thorder}) contains a large number of terms, but if we apply rotating-wave approximations (RWA), the only stationary terms are diagonal terms (which preserves excitation numbers in all modes) and off-diagonal terms that converts excitations between specific modes if certain frequency-matching conditions are satisfied.  In this case, we choose drive frequencies, $\omega_1$ and $\omega_2$, close to the following frequencies
\begin{align}
    &\omega_1 \approx \omega_B + \omega_R - \omega_A\\
    &\omega_2 \approx 2 \omega_A - \omega_R
\end{align}
Under RWA, we have:
\begin{align}
\hat{H} &= \hat{H}_0 + \hat{H}_{SS} +\hat{H}_{kerr} +  \hat{H}_{conv}
\label{eq:H_expanded}\\
\frac{\hat{H}_0+\hat{H}_{SS}}{\hbar} &= (\omega_A-\delta\omega_A) \hat{a}^\dagger \hat{a}+(\omega_B-\delta\omega_B)\hat{b}^\dagger \hat{b}\nonumber\\
&+(\omega_R-\delta\omega_R)\hat{r}^\dagger \hat{r}\\
\frac{\hat{H}_{kerr}}{\hbar} =& -\alpha_{A} \hat{a}^{\dagger 2} \hat{a}^2-\alpha_{B} \hat{b}^{\dagger 2} \hat{b}^2-\alpha_R\hat{r}^{\dagger 2} \hat{r}^2- \chi_{AB} \hat{a}^\dagger \hat{a}\hat{b}^\dagger \hat{b}\nonumber\\
&-\chi_{AR} \hat{a}^\dagger \hat{a}\hat{r}^\dagger \hat{r}-\chi_{BR}\hat{b}^\dagger \hat{b}\hat{r}^\dagger \hat{r}\\
\frac{\hat{H}_{conv}}{\hbar} =& \big(\Omega_1 e^{-i\omega_1 t}\hat{a}\hat{b}^\dagger \hat{r}^\dagger + \Omega_2 e^{-i\omega_2 t} \hat{a}^2 \hat{r}^\dagger + h.c.\big)
\end{align}

%(with small detunings $\delta_1$ and $\delta_2$), and near-stationary 4$^{th}$-order terms of the form $\xi_1\hat{a}\hat{b}^\dagger\hat{r}^\dagger$ and $\xi_2\hat{a}^2\hat{r}^\dagger$ $(+h.~c.)$ emerge as a result of four-wave mixing. We can now rewrite our Hamiltonian as the linear component in addition to all the non-rotating 4th order terms and the two four wave mixing terms.

\begin{figure*}[tbp]
    \centering
    \includegraphics[scale=0.75]{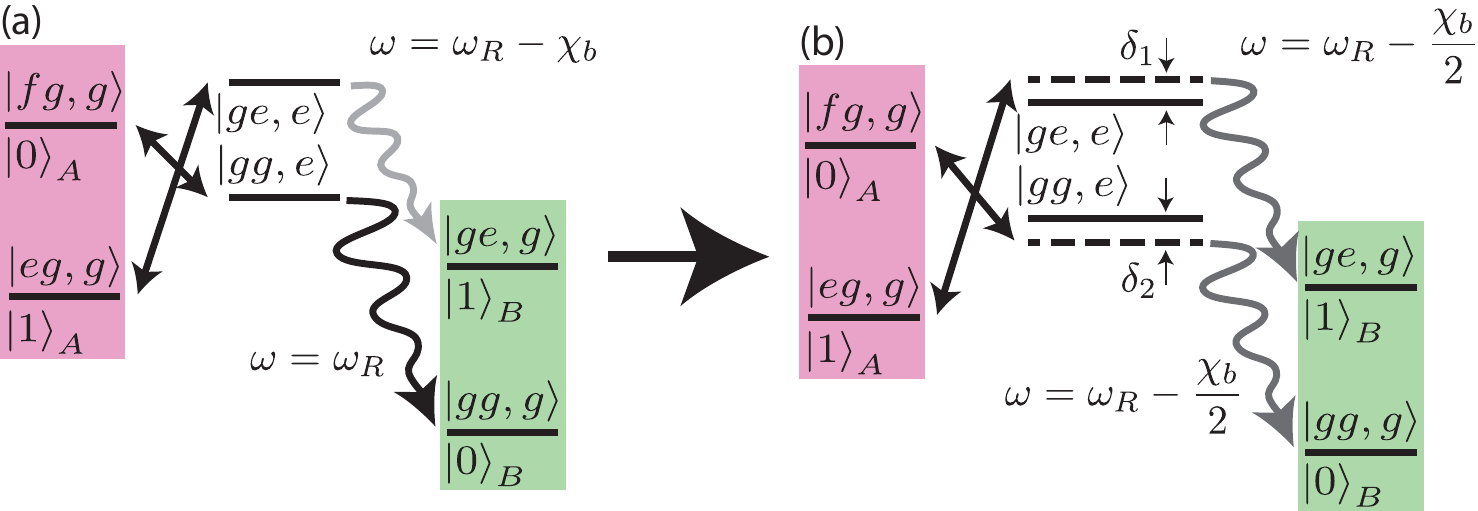}%[width=8.0cm]
    \caption{Comparison of (a) driving resonantly versus (b) driving detuned from resonance through the intermediate states in a circuit QED implementation of AQST. In (a), the emitted photon has frequency $\omega$ or $\omega-\chi_b$ depending on the logical state.  In (b), the emitted photon has identical frequency regardless of logical state when we drive with detunings $\delta_1=-\chi_b$, $\delta_2=\chi_b$.}
    \label{virtual_transfer}
\end{figure*}

%Under the rotating wave approximation (RWA), the Hamiltonian in the reference frame of the drives is
%\begin{align}
%\hat{H}_{\text{rot}}=&\hbar\delta_1 |ge,e\rangle \langle ge,e|+\hbar\delta_2 |gg,e\rangle\langle gg,e| \nonumber\\
%+& \hbar \frac{\Omega_1}{2}|ge,e\rangle\langle eg,g| + \hbar \frac{\Omega_2}{2}|gg,e\rangle\langle fg,g| + h.~c.
%	\label{eq:H_rot}
%\end{align}
Here $\hat{H}_0$ is the original linear contributions of the modes, $\hat{H}_{SS}$ contains the Stark shifts caused by both drives and the Lamb shift caused by junction anharmonicity:
\begin{equation}
\delta\omega_X = \sum_{i=I,II} \frac{E_{Ji}}{\hbar} \big(\Phi^2_{Xi}\Phi^2_{Ri}|\tilde{\xi_1}|^2+\Phi^2_{Xi}\Phi^2_{Ri}|\tilde{\xi_2}|^2+\frac{1}{2}\Phi^4_{Xi}\big)
\end{equation}
$\hat{H}_{kerr}$ contains the anharmonicity (self-Kerr, $\alpha_X$) of the modes and the dispersive shifts (cross-Kerr, $\chi_{XY}$) between the modes:
\begin{align}
    &\alpha_X = \sum_{i=I,II} \frac{E_{Ji}}{2\hbar}\Phi^4_{Xi}\\
    &\chi_{XY} = \sum_{i=I,II} \frac{E_{Ji}}{\hbar}\Phi^2_{Xi}\Phi^2_{Yi}
\end{align}
%We will be referring to $\chi_{BR}$ as $\chi_b$ for convenience. 
$\hat{H}_{conv}$ describes the targeted four wave mixing terms
where the Rabi drive rates 
\begin{align}
    &\Omega_1 = \sum_i \frac{E_{Ji}}{\hbar} \xi_1 \Phi_{Ai}\Phi_{Bi}\Phi_{Ri}^2\\
    &\Omega_2 = \sum_i \frac{E_{Ji}}{2\hbar} \xi_2 \Phi_{Ai}^2\Phi_{Ri}^2
    \label{eq:Omega}
\end{align}
To implement the protocol, $\xi_1$ and $\xi_2$ are chosen to satisfy $\Omega_1=\Omega_2\equiv \Omega$, and $\omega_1$, $\omega_2$ are chosen specifically as:
\begin{align}
    \omega_1 = &\big[(\omega_B-\delta\omega_B) + (\omega_R-\delta\omega_R)-\chi_{BR}\big] \nonumber\\
    &- \big[\omega_A-\delta\omega_A\big] +\delta_1 \\
    \omega_2 = &\big[2(\omega_A-\delta\omega_A)-\alpha_{A})\big] - \big[\omega_R-\delta\omega_R\big] -\delta_2
\end{align}
As shown in Fig.~\ref{virtual_transfer}(b), $\delta_1$ ($\delta_2$) represents a small detuning of the drive tone 1 (2) from the transitions $\ket{eg,g}\rightarrow\ket{ge,e}$ ($\ket{fg,g}\rightarrow\ket{gg,e}$) after accounting for Start shifts.

Now we apply the following transformation to go in a rotating frame:
\begin{align}
\ket{0}_B &\equiv \ket{\widetilde{gg,g}} = \ket{gg,g}
\label{eq:rotatingframe1}\\
\ket{1}_B &\equiv \ket{\widetilde{ge,g}} = e^{-i \omega_B}\ket{ge,g}\\
\ket{0}_A &\equiv \ket{\widetilde{fg,g}} = e^{-i (2\omega_A - \alpha_{A})}\ket{fg,g}\\
\ket{1}_A &\equiv \ket{\widetilde{eg,g}} = e^{-i \omega_A}\ket{eg,g}\\
\ket{0}_e &\equiv \ket{\widetilde{gg,e}} = e^{-i (\omega_R - \delta_2)}\ket{gg,e}\\
\ket{1}_e &\equiv \ket{\widetilde{ge,e}} = e^{-i (\omega_R + \omega_B - \chi_b - \delta_1)}\ket{ge,e}
\label{eq:rotatingframe6}
%\\\ket{\widetilde{eg,e}} = e^{-i (\omega_R + \omega_A - \chi_B /2)}\ket{eg,e}\\
%\ket{\widetilde{fg,e}} = e^{-i (\omega_R  + 2\omega_A - \chi_A + \chi_B /2)}\ket{fg,e}
\end{align}
where $\chi_b\equiv\chi_{BR}$ is the simplified notation used in the main text for the dispersive shift between B and R.  The system Hamiltonian Eq.~(\ref{eq:H_expanded}), within the Hilbert space of the 6 relevant states, is transformed to:
\begin{align}
\frac{\hat{H}}{\hbar} = \delta_1 \ket{\widetilde{ge,e}} \bra{\widetilde{ge,e}}+\delta_2 \ket{\widetilde{gg,e}}\bra{\widetilde{gg,e}} \nonumber\\
+ \Omega \big( \ket{\widetilde{ge,e}}\bra{\widetilde{eg,g}} + \ket{\widetilde{gg,e}}\langle\widetilde{fg,g}|  + h.c. \big)
%\frac{\hat{H}}{\hbar} = (\chi_{b}+\delta_1) \ket{1}_e \bra{1}_e+\delta_2 \ket{\widetilde{gg,e}}\bra{\widetilde{gg,e}} + \Omega \big( \ket{\widetilde{ge,e}}\bra{\widetilde{eg,g}} + \ket{\widetilde{gg,e}}\langle\widetilde{fg,g}|  + h.c. \big)
\label{eq:H_rot}
\end{align}
The reservoir loss operator $\hat{L}=\sqrt{\kappa}\hat{r}$ is transformed to Eq.~(13):
\begin{equation}
\hat{L}_{\text{rot}}=\sqrt{\kappa}\big(|\widetilde{gg,g}\rangle\langle \widetilde{gg,e}|+ e^{i(\delta_2-\delta_1-\chi_{b}
)t}|\widetilde{ge,g}\rangle\langle \widetilde{ge,e}|\big)
\end{equation}
The-time dependent phase factor in $\hat{L}_{\text{rot}}$ indicates a dephasing effect due to the energy difference of the reservoir emission for logical $\ket{0}$ versus $\ket{1}$.  To eliminate this error, we may choose detunings $\delta_1=-\chi_b/2$ and 
$\delta_2=\chi_b/2$ to make $\hat{L}_\text{rot}$ stationary:
\begin{align}
\hat{L}_{\text{rot}}=\sqrt{\kappa}\big(|\widetilde{gg,g}\rangle\langle \widetilde{gg,e}|+|\widetilde{ge,g}\rangle\langle \widetilde{ge,e}|\big)
\label{eq:final_loss}
\end{align}
Effectively, we drive the two sets of transitions through nearby virtual states to compensate for the dispersive shift of the real states.  These symmetrically chosen detunings also ensure equal rates ($=\sqrt{\Omega^2+(\chi_b/2)^2}$) for the two detuned Rabi drives.  In the main text and in the discussions below, we refer to all states in the rotating frame directly omitting the use of the "tilde" signs. 

%The different rotation axes of the two detuned Rabi drives make the two $\Lambda$-transitions still distinguishable by the environment, but the resulting infidelity is small in the limit of $\chi_b\ll\kappa$ as a result of the Zeno effect (scales as $\chi_b^2/\kappa^2$ and independent of $\Omega$ as we find numerically, Fig.~2(c) inset).

\section{Intrinsic Infidelity from Detuned Drives in cQED Implementation}
The most straightforward implementation of AQST in this proposed cQED device is to populate the two intermediate states ($\ket{0}_e\equiv\ket{gg,e}$, $\ket{1}_e\equiv\ket{ge,e}$) with on-resonance drives ($\delta_1=\delta_2=0$).  However, because these two intermediate states have an energy difference unequal to the difference between the final states ($\ket{0}_B\equiv\ket{gg,g}$, $\ket{1}_B\equiv\ket{ge,g}$) by $\chi_b$, there is an intrinsic infidelity introduced to the transfer scheme. In this case, the transfer of the logical $\ket{0}$ or logical $\ket{1}$ state emits a photon of different frequency (Fig.~\ref{virtual_transfer}(a)), and the environment receiving the photon may acquire this ``which-frequency" information and hence collapse the quantum state.  The infidelity of the transfer, governed by the Heisenberg uncertainty of the frequency of the emitted photon, depends on the time over which the transfer occurs.  Quick transfer will have a relatively small time uncertainty and a relatively large frequency uncertainty that limits the information leakage to the environment. In practice, we expect that the transfer speed will be limited by the relatively small four wave mixing transition rate $\Omega$. When $\Omega\ll\kappa$, this will bound the total transfer rate to $\kappa_\text{eff} \approx 4 \Omega^2 / \kappa$, and high-fidelity transfer will require $\chi_b\ll\kappa_\text{eff}$.  In this limit the transfer infidelity will be proportional to $(\chi_b/\kappa_\text{eff})^2$, but it is challenging to approach this limit.  

%\begin{figure}[tbp]
%    \centering
%    \includegraphics[scale=1]{SM2.pdf}%[width=8.0cm]
%    \caption{}
%\end{figure}

Instead we drive with detunings of $\delta_1 = -\chi_b /2$ and $\delta_2 = \chi_b / 2$ to make the frequencies of these two decay transitions equal. This makes the loss operator $\hat{L}$ stationary so that the environment no longer decoheres the transferred state by discerning the frequency of the emitted photon (Fig.~\ref{virtual_transfer}(b)). However, this does introduce a new error due to the fact that excitations in the intermediate states are now rotating in the drive frame (with angular frequency of $+\chi_b /2$ or $-\chi_b / 2$) and accumulating different phases depending on the logical qubit state.  Therefore, there will a relative phase imprinted between logical $\ket{0}$ and $\ket{1}$ through the state transfer process.  Furthermore, this phase depends on the random timing of the reservoir dissipation process and causes decoherence.  This intrinsic infidelity turns out to have a more favorable scaling: $1-\mathcal{F} = \frac{1}{2}(\chi_b/\kappa)^2$ for an equator state (which is the worst case), as we found by numerical simulation of the master equation (without considering experimental non-ideality such as $T_1$, $T_2$ processes, Fig.~3(c)).  We will discuss the limiting case of $\chi_b\ll\Omega\ll\kappa$ to shed light on this scaling and why it is independent of $\Omega$. 

\begin{figure*}[tbp]
    \centering
    \includegraphics[scale=0.6]{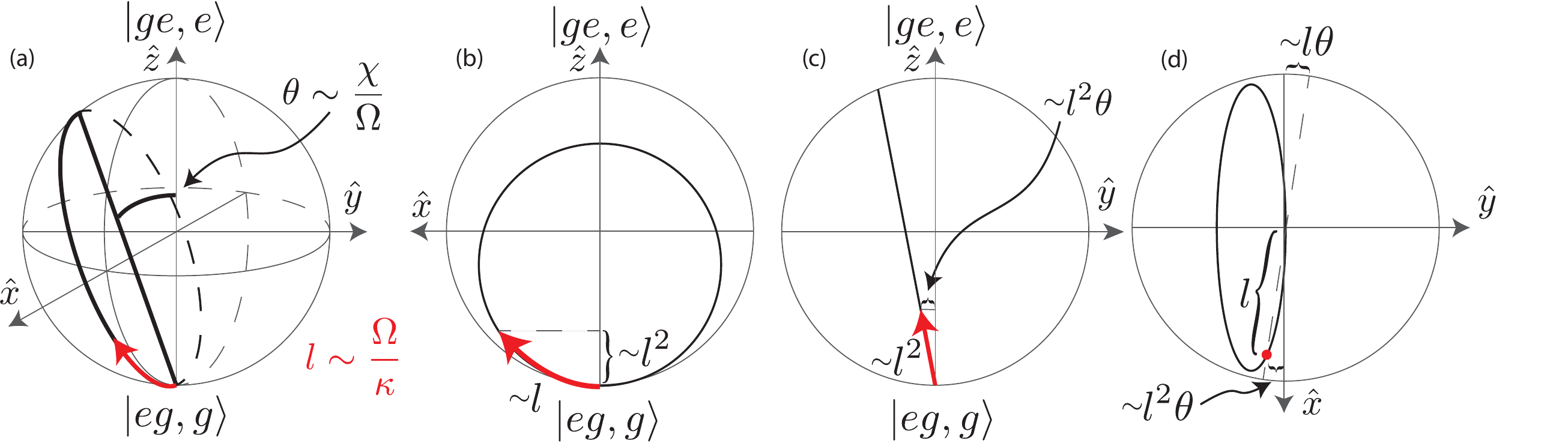}%[width=8.0cm]
    \caption{Schematic visualization of the trajectory of a logical $\ket{0}$ during AQST in the limit of $\chi_b\ll\Omega\ll\kappa$, within the Bloch sphere of $\ket{eg,g}$ and $\ket{ge,e}$, including (a) 3D view, (b) projection in x-z plane, (c) projection in y-z plane, (d) projection in x-y plane.
    }
    \label{Detuned-Rabi}
\end{figure*}

The scaling of this infidelity can be intuitively considered via the trajectory for a logical $\ket{1}$ state during the transfer in the rotating frame of the drives (Fig.~\ref{Detuned-Rabi}).  Consider the textbook picture of Rabi rotation under a slightly-detuned drive, where the quantum state is driven from the south pole of the Bloch sphere ($\ket{eg,g}$) towards near the north pole ($\ket{ge,e}$) along a circle slightly tilted from the x-z plane by an angle of $\theta = \delta/2\Omega = \chi/4\Omega$.  When the quantum state is driven out of the south pole, it is subject to reservoir dissipation which suppresses its dynamics (Zeno effect) and eventually completes the AQST via a quantum jump out of this Bloch sphere.  The distance that the quantum state travels along the Bloch sphere (assumed with radius of 1) can be represented by the red arc, whose length $l$ is of the order $\Omega/\kappa \ll 1$.  This arc's projected height along z-axis is of the order $l^2$ (Fig.~\ref{Detuned-Rabi}(b)), and its projection along the y-axis is therefore of the order $l^2\theta$ (Fig.~\ref{Detuned-Rabi}(c)).  The accumulated phase of the transferred logical $\ket{1}$ state relative to the original state is given by the azimuth angle of the trajectory, which is $l\theta$ or on the order of $\chi_b/\kappa$.  Note the cancellation of $\Omega$ here.  Since the trajectory for a logical $\ket{0}$ in its own Bloch sphere is mirror symmetric to Fig.~\ref{Detuned-Rabi}, there will be an added relative phase between the two logical states of order $\chi_b/\kappa$, giving infidelity of order $(\chi_b/\kappa)^2$.

More rigorous calculation of this intrinsic infidelity can be done analytically using the effective Hamiltonian approach similar to Appendix A.  Before quantum jump occurs, the stochastic wavefunction follows the Schrodinger equation:
\begin{equation}
    i\hbar\frac{d\ket{\phi(t)}}{dt}=\hat{H}_\text{eff}\ket{\phi(t)}
    \label{eq:Schrodinger2}
\end{equation}
with non-Hermitian effective Hamiltonian (from Eq.~(\ref{eq:H_rot})):
\begin{align}
\frac{\hat{H}_\text{eff}}{\hbar} =& \frac{\hat{H}}{\hbar} - \frac{i}{2}\hat{L}^\dagger_\text{rot}\hat{L}_\text{rot} \nonumber\\
=& \frac{\kappa-i\chi_b}{2i} \ket{1}_e \bra{1}_e+\frac{\kappa+i\chi_b}{2i} \ket{0}_e\bra{0}_e \nonumber\\
&+ \Omega \big( \ket{1}_e\bra{1}_A + \ket{0}_e\langle{0}|_A  + h.c. \big)
\end{align}
Solving the no-jump dynamics in the limit of $\Omega\ll\kappa$, the system wavefunction, starting from an equator state $\ket{\phi}_\text{init}=(\ket{0}_A+\ket{1}_A)/\sqrt{2}$, will first quickly converge to a quasi-steady (dark) state over a time scale of $1/\kappa$. To first order in $\Omega/\kappa$, %for $t \ll \kappa/(4\Omega^2)$:
\begin{align}
    &\ket{\phi(t)} = \frac{1}{\sqrt{2}}\big(\ket{0}_A + \ket{1}_A \big) - \frac{1}{\sqrt{2}}\bigg\{\frac{2i\Omega}{\kappa-i\chi_b} \nonumber\\
    &\big[1-e^{-\frac{1}{2}(\kappa-i\chi_b)t}\big]\ket{0}_e 
    + \frac{2i\Omega}{\kappa+i\chi_b}\big[1-e^{-\frac{1}{2}(\kappa+i\chi_b)t}\big]\ket{1}_e\bigg\}
    \label{eq:short-dynamics}
\end{align}
Noting that the $\ket{0}_e$ and $\ket{1}_e$ components have phases of $\pm\chi_b/\kappa$ respectively at this converged quasi-steady state ($1/\kappa \ll t \ll \kappa/\Omega^2$).  If a quantum jump were to occur at this quasi-steady state, the final state would be:
\begin{equation}
    \ket{\phi}_\text{final}=\frac{1}{\sqrt{2}}\big(e^{i\eta}\ket{0}_B + e^{-i\eta}\ket{1}_B\big)
\end{equation}
with $\eta = \chi_b/\kappa$.  For $\chi_b\ll\kappa$, This would give an infidelity of $\eta^2 = (\chi_b/\kappa)^2$ relative to the target final state of $\ket{\phi}_\text{target}=(\ket{0}_B+\ket{1}_B)/\sqrt{2}$.  However, once reaching this quasi-steady state, the wavefunction also rotates slowly over long time scale following:
\begin{align}
    \ket{\phi(t)} = \frac{1}{\sqrt{2}}\bigg(\ket{0}_A -\frac{2i\Omega}{\kappa-i\chi_b}\ket{0}_e \bigg)e^{-i\frac{2\Omega^2\chi_b}{\kappa^2}t} \nonumber\\
    + \frac{1}{\sqrt{2}}\bigg(\ket{1}_A -\frac{2i\Omega}{\kappa+i\chi_b}\ket{1}_e\bigg) e^{i\frac{2\Omega^2\chi_b}{\kappa^2}t}
    \label{eq:long-dynamics}
\end{align}
Therefore, over this long time scale that AQST may occur (whose rate is $4\Omega^2/\kappa$), the final state will acquire a phase angle $\eta$ depending on the exact timing $t$ of the quantum jump:
\begin{equation}
    \eta(t) = \frac{\chi_b}{\kappa} - \frac{2\Omega^2\chi_b}{\kappa^2}t
\end{equation}
In the end, the intrinstic infidelity of the transferred state against the target state $\ket{\phi}_\text{target}$ is an integral weighted by the jump probability $P_\text{jump}(t)$:
\begin{align}
    1-\mathcal{F} &\approx \int_{t=0}^\infty \eta^2 P_\text{jump} dt \nonumber\\
    &=  \int_{t=0}^\infty  \bigg(\frac{\chi_b}{\kappa} - \frac{2\Omega^2\chi_b}{\kappa^2}t \bigg)^2  
    \bigg(\frac{4\Omega^2}{\kappa}\bigg) e^{-\frac{4\Omega^2 t}{\kappa}} dt 
%    \nonumber\\ &=  \int_{\tau=0}^\infty  \bigg(\frac{\chi_b}{\kappa} - \frac{\chi_b}{2\kappa}\tau \bigg)^2 e^{-\tau t} d\tau
    = \frac{\chi_b^2}{2\kappa^2}
    \label{eq:IntInfidelity}
\end{align}

It should be noted that so far we have naively chosen the target final state to be not rotated from the initial state.  However, on average, the phase angle $\eta$ acquired during this AQST process is non-zero:
\begin{equation}
    \langle\eta(t)\rangle = \int_{t=0}^\infty \eta P_\text{jump} dt = \frac{\chi_b}{2\kappa}
\end{equation}
If we choose this average case as the target state: $\ket{\phi'}_\text{target}=(e^{i\chi_b/2\kappa}\ket{0}_B+e^{-i\chi_b/2\kappa}\ket{1}_B)/\sqrt{2}$ (which is indeed the nearest pure-state for the final density matrix), we get an updated infidelity for the transferred state:
\begin{equation}
    1-\mathcal{F'} = \frac{\chi_b^2}{4\kappa^2}
\end{equation}
In essence, our computed infidelity in Eq.~(\ref{eq:IntInfidelity}) and quoted in the main text includes two equal contributions, half being a ``coherent error", or a unitary rotation that in principle can be corrected for, and half being a more intrinsic ``incoherent error" due to the random timing of the quantum jump. 

%The above analytical solution of wave function dynamics in this subsection requires no assumption of $\chi_b\ll\kappa$.  Furthermore, 
When $\Omega\ll\kappa$ is no longer satisfied, even though we can no longer separate the system dynamics into short-time (Eq.~(\ref{eq:short-dynamics})) and long-time (Eq.~(\ref{eq:long-dynamics})) behavior, through master equation simulation, we found the fidelity in Eq.~(\ref{eq:IntInfidelity}) still approximately holds for a large range of the parameters.  This includes the parameter sets of $\chi\approx\Omega\approx\kappa/5$ that we simulated for realistic experiments.

\section{cQED Simulation Parameters and Discussions}

\begin{table}[t]
\begin{tabular}{c c c} % centered columns (4 columns)
\hline\hline\\[-2ex]
  & \,\,\,\,Junction I & Junction II \\
\hline\\[-2ex]
$E_J/2\pi$ & \,\,\,\,40 GHz        & 56 GHz    \\
$\Phi_A$ & \,\,\,\,0.03        & 0.23         \\ 
$\Phi_B$ & \,\,\,\,0.0025-0.0141        & 0.002        \\
$\Phi_R$ & \,\,\,\,0.32        & 0.01         \\ [0.5ex]
% [1ex] adds vertical space
\hline
\end{tabular}
\caption{simulation parameters: Josephson energy of Junctions I and II, and the zero point fluctuations (ZPF) across each of them due to excitations in modes A, B, and R. ZPF of $B$ mode across junction I is a variable parameter in the simulation.}
\label{tab:my-table}
\end{table}
Simulations of the cQED implementation were performed with Qutip, a python-based master equation solver. This allows us to simulate both Hamiltonian terms and any potential jump/loss operator. Our system is composed of a 3-level system ($A$), and two 2-level systems ($B$ and $R$) making a 12-level Hilbert space. We simulate under the rotating wave approximation in the rotating frame defined by Eq.~(\ref{eq:rotatingframe1})-(\ref{eq:rotatingframe6}), canceling out all bare mode energies. Initial states are encoded in a pure superposition of $\ket{fg,g}$ and $\ket{eg,g}$. The initial states are coupled to the states $\ket{gg,e}$ and $\ket{ge,e}$ through the four wave mixing (FWM) process, simulated through off-diagonal Hamiltonian terms with amplitude $\Omega$, see Eq.~(\ref{eq:H_rot}). %This is simulated, through an off-diagonal Hamiltonian term, rather than a loss operator. 
These two intermediary states are slightly detuned from the bare states by $\pm\chi_b$%and $\delta_2$
, as reflected by appropriate diagonal elements for these two states (Eq.~(\ref{eq:H_rot})) and other non-computational states. The loss acting upon the reservoir mode is then stationary with respect to the intermediate states and can be simulated with a time-independent operator $\hat{L}$ as in Eq.~(\ref{eq:final_loss}).

\begin{table*}[bt]
\caption{Frequencies, nonlinear couplings ($\chi$ matrix $/2\pi$), and relaxation times of modes $A$, $B$ and $R$ used in our simulation.  For the $\chi$ matrix, diagonal terms are the mode anharmonicities and off-diagonal terms are dispersive frequency shifts. $\chi_{BR}$ and the reservoir lifetime (= $2\pi/\kappa$) are variable parameters in the simulation, with the range of values for the graph in Fig.~3 reported here.  The loaded $T_1$ time of $A$ and $B$ (including the Purcell effect) are varied accordingly.  The $T_1$ time of $\ket{f}_A$ is assumed to be half of that of $\ket{e}_A$ (which is listed here).  *The frequency of cavity B can be chosen arbitrarily with no effect to the simulation.}
\centering  
\begin{tabular}{c c c c c c c } % centered columns (4 columns)
\hline\hline\\[-2ex]
		& Frequency	&	 &Nonlinear coupling&\,\,\,\,\,\,\,\,&Intrinsic &Loaded\\
		& $\omega/2\pi$	& A		& B	& R & $T_1$ & $T_1$\\
\hline\\[-2ex]
$A$\,\,\,	& 5.9 GHz	& 78 MHz	&  	        &	        &50 $\mu$s  & 14-42 $\mu$s \\
$B$\,\,\,	& \,\,\,\,6.5 GHz*\,\,\,\,		& \,0.01 MHz\,\,	&  0-0.8 kHz	&	        &800 $\mu$s & 80-500 $\mu$s\\
$R$\,\,\,	& 8.0 GHz		& 4.0 MHz	&  0.03-0.82 MHz&	210 MHz && 0.05-0.7 $\mu$s\\[0.5ex]
% [1ex] adds vertical space
\hline
\end{tabular}
\label{tab:my-table2}
\end{table*}

In order to faithfully assess the potential of this scheme in cQED, we must consider a number of errors that could spoil the process fidelity. The errors here are simulated with loss operators: 1) Both $A$ and $B$ have intrinsic loss as well as Purcell loss introduced by couplings to $R$. 2) Loss out of the second excited state in $A$ is simulated with twice the rate as loss from the first excited state. 3) The reservoir can be excited by a hot environment due to non-ideal thermalization, which is simulated with a rate $\Gamma_{\uparrow}$. This is expected to be by far the dominant dephasing mechanism in $A$ and $B$. 

Simulation parameters were selected to be realistic with transmons in a coaxial 3D cavity architecture~\cite{reagor_quantum_2016}. For all simulations we start with the two junction energies ($E_{Ji}$) and the zero point fluctuation (ZPFs) across each junction from each mode ($\Phi_{Xi}$) as shown in Table \ref{tab:my-table}. %, and the rates of our loss operators. 
Junction energies and the ZPFs, except $\Phi_{BI}$, are kept constant throughout all simulations. These parameters uniquely define the mode frequencies, anharmonicities and dispersive couplings as shown in Table \ref{tab:my-table2} (with the exception of $\omega_B$) assuming $A$ and $R$ are transmon-like modes with no additional linear inductance.  (If desirable, $\omega_A$ and $\omega_R$ can be reduced without impacting any other Hamiltonian terms by introducing linear inductance to them.)  % From these we calculate frequencies and couplings between modes. 
Frequencies of all objects were kept in the 4-8 GHz range, convenient for most experimental setups, but they play no explicit role in the simulation because of the rotating wave approximation. %$A$ and reservoir frequencies are derivable by the junction ZPFs and the junction energies but the $B$ frequency is totally independent of our simulation parameters. We choose 6GHz to achieve reasonable frequencies of our FWM drives. 
%The coupling between $R$ and $B$ is critical to the simulations and is chosen to be slower than the reservoir relaxation and faster than the FWM drives. 

For the plot in Fig.~3(c), we sweep the dispersive coupling between $B$ and $R$, $\chi_b$, by changing $\Phi_{BI}$. For each $\chi_b$, because the relaxation rate of the reservoir can be engineered at will, it is always beneficial to maximize $\xi_{1,2}$ and hence the FWM rate $\Omega$ to the extent possible.  $\xi$ is limited in practice by heating effects due to higher-order Josephson non-linearity~\cite{zhang_engineering_2019}, and we consider a conservative upper bound of $\xi_{max} = 0.3$, compared with $\xi\approx0.35$ in Ref.~\cite{gao_programmable_2018} and $\xi\approx0.5$ in Ref.~\cite{rosenblum_cnot_2018}.  
%, we attempt to maximize our FWM rate. 
Because the linear cavity $B$ has the relatively small junction ZPFs ($\Phi_{BI}$ and $\Phi_{BII}$), larger $\xi_1$ than $\xi_2$ is needed to achieve $\Omega_1=\Omega_2$, as indicated by Eq.~(\ref{eq:Omega}). %  to  so for all simulations $\Omega_2 > \Omega_1$. For the protocol, we require $\Omega_1=\Omega_2$ 
Therefore, we always maximize $\Omega_1$ by maximizing $\xi_1$ and then choose $\xi_2$ accordingly, \textit{i.e.}~$\xi_2<\xi_1=0.3$.  For each $\chi_b$, after maximizing $\Omega$, we then sweep over values of $\kappa$ to maximize the state transfer fidelity. Increasing $\chi_b$ increases the transfer speed but has drawbacks we need to consider. Larger $\chi_b$ causes an increase in Purcell loss through the reservoir, an increase in dephasing from thermal shot noise of the reservoir and an increase in intrinsic infidelity as discussed in the last subsection.  We report overall state transfer fidelity as an average of the fidelities of the six cardinal points on the logical Bloch sphere (between $\ket{0}_L$ and $\ket{1}_L$). %Each of the fidelity plots shown are for an equator state.

For the inset in Fig.~3(c), no other forms of decoherence or relaxation other than the reservoir $\kappa$ were implemented, to demonstrate the scaling of the intrinsic infidelity of the proposed protocol with the ratio of $\kappa / \chi_B$. Each point is the measure of the infidelity as $t \rightarrow \infty$ while varying $\chi_B$ for each simulation. We found this infidelity to scale as $(\chi_b/\kappa)^2/2$ over a very large parameter range even when $\chi_b\ll\Omega\ll\kappa$ is no longer satisfied. 

The transfer scheme requires a 3 $\times$ 2 system in addition to the reservoir mode, but only half of these states are inside the logical space of the transfer. States outside the useful computational space include $\ket{eg,e}, \ket{ee,g}, \ket{ee,e}, \ket{fg,e}, \ket{fe,g}$, and $\ket{fe,e}$. For these states only diagonal Hamiltonian terms are present. Notably, none of these states are the result of any loss operator acting on any of the computational states, so energy relaxation will not bring the system outside of the logical space.  Because of its short $T_1$ time, for typical non-ideal thermalization the reservoir will have a significant $\Gamma_\uparrow$ rate, which can excite the system to $\ket{fg,e}$ and $\ket{eg,e}$ (and quickly relax back).  This is accounted for in the simulation by considering a rate $\Gamma_{up}$ to be $\kappa/100$, corresponding to a thermal population of 1\% for the reservoir, comparable to reservoir modes strongly coupled to the transmission line in Ref.~\cite{wang_schrodinger_2016, touzard_coherent_2018}.  This effectively captures the dominant dephasing effects in both $A$ and $B$, and we do not consider any additional dephasing that $A$ and $B$ may experience.  This is because the internal dephasing rate for fix-frequency transmons or linear cavities, if there is any, is much smaller than other error rates in our simulation, and dephasing due to other peripheral (\textit{i.e.}~readout) modes can also be minimized by choosing relatively slow rates for them.  

We do not account for a $\Gamma_{\uparrow}$ in $A$ or $B$ as we expect them to be negligibly small excluding the possibility of accessing the states $\ket{fe,g}, \ket{ee,g}, \ket{ee,e}$ and $\ket{fe,e}$.  In addition, we have neglected the leakage error out of the 12-dimension Hilbert space is not included, but its leading contribution from spurious transition of $R$ to its second excited state is estimated to be less than 0.2\%.

\end{appendix}

\bibliographystyle{apsrev4-1}
\bibliography{Zotero}

\end{document}